\documentclass{aa}
\usepackage{graphicx}
\usepackage{txfonts}

\begin{document}

%%-----------------------------
%%      the top matter
%%-----------------------------

\title{Buoyancy-induced time delays in Babcock-Leighton flux-transport
  dynamo models}

\author{Laur\`ene Jouve \and Michael R. E. Proctor \and Geoffroy Lesur}

\offprints{Jouve Laur\`ene}

\institute{DAMTP, Wilberforce Road, CB3 0WA Cambridge, UK\\
              \email{lj272@damtp.cam.ac.uk}}

%\author{Author3, C.}\address{...}
%

%
%\setcounter{page}{1}

%
\abstract
{The Sun is a magnetic star whose cyclic activity is thought to be
  linked to internal dynamo mechanisms. A combination of numerical
  modelling with various levels of complexity is an efficient and accurate tool to investigate such
  intricate dynamical processes. }
{We investigate the role of the magnetic buoyancy process in
  2D Babcock-Leighton dynamo models, by modelling more accurately the
  surface source term for poloidal field.}{ To do so, we reintroduce in
  mean-field models the results of full 3D MHD calculations of the
  non-linear evolution of a rising flux tube in a convective
  shell. More specifically, the Babcock-Leighton source term is
  modified to take into account the delay introduced by the rise time
  of the toroidal structures from the base of the convection zone to
  the solar surface.}  
  {We find that the time delays introduced in the equations produce
  large temporal modulation of the cycle amplitude even when strong
  and thus rapidly rising flux tubes are considered. Aperiodic
  modulations of the solar cycle appear after a sequence of period
  doubling bifurcations typical of non-linear systems. The strong
  effects introduced even by small delays is found to be due to the
  dependence of the delays on the magnetic field strength at the base
  of the convection zone, the modulation being much less when time delays remain constant. We do not find
  any significant influence on the cycle period except when the delays are made
  artificially strong.}{ A possible new origin of the solar cycle variability
  is here revealed. This modulated activity and the resulting
  butterfly diagram are then more compatible with observations than
  what the standard Babcock-Leighton model produces.}

 
\authorrunning{Jouve, Proctor \& Lesur}
\titlerunning{Time delays and BL dynamo models}

\maketitle

%
%%-----------------------------
%%      your text
%%-----------------------------
\section{Introduction}

Our Sun is a prime example of a very turbulent and magnetically active
star. Its robust 22-yr activity cycle originates in the periodic polarity
reversal of the Sun's internal large-scale magnetic field. As a
result, sunspots emerge at the solar surface with statistically well-defined dynamical and morphological characteristics. These sunspots
are thought to be the surface manifestations of strong toroidal
magnetic structures created at the base of the convection
zone and which rise through the plasma under the effect of magnetic buoyancy (\cite{Parker55}). This toroidal field undergoes cyclic variations along with the
poloidal field, which flips polarity at sunspot cycle maxima. Even if
a robust regular activity is easily exhibited in the Sun, a significant
modulation of both the amplitude and the frequency of the cycle has
been observed. In particular, periods of strongly reduced activity
have been
revealed, the most famous being the Maunder minimum between 1650 and
1700, during when no sunspots were to be seen at the solar surface
(\cite{Eddy76}). Today, understanding the origins of such a
variability has become crucial. Not only is the impact on satellites,
astronauts, power grids or radio communications significant but the
climatologists community now tends to address more and more the
question of the
influence of solar variability on the Earth climate
(e.g. \cite{Bard06}, \cite{Lean05}).

The classical explanation for the cyclic activity of the
large-scale magnetic field is that a dynamo process acts in the solar
interior to regenerate the three components of the magnetic field and
sustain them against ohmic dissipation. The inductive action of the
complex fluid motions would thus be responsible for the vigorous
regeneration of magnetic fields and for its non-linear evolution in
the solar interior (see \cite{Charbonneau051} and \cite{Miesch05} for
recent reviews on the subject).

Understanding how these complex physical processes operating in the solar
turbulent plasma non-linearly interact is very challenging. One
successful and powerful approach is to rely on multi-dimensional
magnetohydrodynamic (MHD) simulations. In this context, two types of
numerical experiments have been performed since the 70's: kinematic
mean-field axisymmetric dynamo models which solve only the mean
induction equation (\cite{Steenbeck69, Roberts72, Stix76, Krause80}) and
full 3D global models which explicitly solve the full set of
MHD equations (\cite{Gilman83, Glatzmaier85,
  Cattaneo99, Brun04}). Those two approaches are complementary
  and needed since 2D mean-field models are limited by the fact that
  they rely on simplified descriptions of complex physical processes
  such as turbulence and since the cost of 3D models make it
  difficult, as of today, to provide any reliable predictions concerning
  the large-scale magnetic cycle.

As far as the first kind of numerical simulations is concerned, a
particular model has been favoured by a part of the community, namely the Babcock-Leighton
flux-transport model first proposed by \cite{Babcock61} and
\cite{Leighton69}. This model has proved to be very efficient at
reproducing several properties of the solar cycle as the equatorward
branch of activity or the phase relationship between toroidal and
poloidal fields. It has thus been extensively used since the 90's
(\cite{Wang91, Durney95, Dikpati99, Nandy01}),
even to make tentative predictions of the next solar cycle
(\cite{Dikpati06}). In this formulation, the poloidal field
owes its origin to the tilt of active regions emerging at the
Sun's surface at 
various latitudes during the solar cycle. It thus relies on a
  different mechanism than ``classical'' $\alpha\Omega$ dynamo models for
  which the poloidal field is generated by turbulent helical
  motions twisting and shearing toroidal field lines inside the
  convection zone. The emergence of tilted
active regions at the photosphere is the result of the complex non-linear evolution of
strong toroidal structures from the base of the convection zone (CZ), where
they are created through the $\Omega$-effect. It is
thus natural to rely on 3D MHD simulations of rising toroidal structures
in the convection zone to gain some insight on the best way to model
the Babcock-Leighton source term.

 Many models carried out since the 80's relied
on the assumption that toroidal flux is organised in the form of
discrete flux tubes which will rise cohesively from the base of the CZ
up to the solar surface. These models enabled to
demonstrate that the initial strength of magnetic field was an
important parameter in the evolution of the tube and that the active
regions tilts could be explained by the action of the Coriolis force
on the magnetic structure (\cite{DSilva93, Fan94, Caligari95}).
More sophisticated
multidimensional models were then developed and extended to the upper
 part of the CZ and the transition to the solar atmosphere
 (\cite{Magara04, Archontis05, Cheung07,  Martinez08}).
Computations were then performed to study the influence of convective
turbulent flows on the dynamical evolution of flux ropes inside the
CZ (\cite{Dorch01, Cline03, Fan03}, Jouve $\&$
Brun 2009). In particular in \cite{Jouve09}, such computations
were made in a global spherical geometry and thus allowed to assess the
combined role of 
hoop stresses, Coriolis force, convective plumes, turbulence, mean flows and sphericity 
on the tube evolution and on the subsequent emerging regions, along with the usual parameters such as field strength,
 twist of the field lines or magnetic diffusion.
How can the results of 3D simulations of this presumed major step of
the dynamo loop be reintroduced in 2D mean-field models? This is the question we
are addressing in this work. A particular point we retain from these
simulations is that the rise velocity
and thus the rise time of magnetic structures is strongly dependent on the
initial field strength at the base of the convection
zone. Indeed, magnetic buoyancy competes with convective flows to
control the trajectory and the speed of the flux tubes while they rise. Since the Babcock-Leighton source term relies on this process to
regenerate poloidal fields, it may be worth trying to introduce magnetic-field dependent time delays caused by rising toroidal fields
in those models and analyze the influence on the solar cycle.

Indeed, time delays built up in solar
dynamo models have
been shown to cause long-term modulation of the dynamo cycle and under
some circumstances to lead to a chaotic behaviour (\cite{Yoshimura78},
\cite{Charbonneau052} or \cite{Wilmot05}).
In the framework of mean-field dynamo theory, several possibilities
have been studied to explain the variability of the solar cycle. They
mainly fell into two categories: stochastic forcing or dynamical
nonlinearities. Indeed, as we mentioned above, the solar convection zone is
highly turbulent and it would be surprising if the dynamo processes
acting inside this turbulent plasma were nicely regular. The influence
of stochastic
fluctuations in the mean-field dynamo coefficients has been studied in
various models (e.g. \cite{Hoyng88,Ossendrijver96,Weiss00,Charbonneau00}). Moreover, the dynamical
feedback of the strong dynamo-generated magnetic fields is likely to
be significant enough to produce non-linear effects on the activity
cycle. A number of models have introduced these non-linear effects
(\cite{Proctor77, Tobias97, Moss00,Bushby06,Rempel06}) and
have resulted in the production of grand minima-like periods or other
strong modulation of the cyclic activity.  
However, time delays have hardly been considered and if they were,
they were mainly due to the advection time by meridional flow. Indeed, the time-scale of the buoyant
rise of flux tubes was considered to be so small compared to the cycle
period (and to the meridional flow turnover time) that this particular
step was assumed to be instantaneous. However, we would like to
address the question of the influence of magnetic field dependent
delays on the cycle produced by Babcock-Leighton models and especially
on its potential modulation. We propose to do so in the present paper.

This article is organized as follows: Section \ref{sect_3D} summarizes
the results of 3D calculations which will be used in our modified 2D mean-field
Babcock-Leighton model. The formulation of this new model is then
presented, with a particular focus on the modification of the surface
source term. Results of 2D models are shown in Sect. \ref{sect_2D} and
analyzed in the following two sections. Sect. \ref{sect_6th} and
\ref{sect_5th} present and study the behaviour of
a reduced set of ordinary differential equations designed to gain some
insight on the results of the 2D model. We discuss the results and
conclude in Sect. \ref{sect_conclu}.

\section{Summary of the 3D calculations}
\label{sect_3D}

In this section, we briefly summarize the results obtained by
\cite{Jouve09}. They use 3D simulations in a rotating spherical convective
shell to compute the non-linear evolution of a magnetic flux tube
inside the turbulent convection zone.

\subsection{Model and background}

The computations presented in \cite{Jouve09} use the ASH code which
solves the anelastic equations of magnetohydrodynamics in three
dimensions and in spherical geometry (\cite{Clune99,Miesch00,Brun04}). The first step of these
calculations is to trigger the convective instability in the rotating
sphere and let the model evolve until it reaches a statistically
steady state. A twisted magnetic flux tube is then introduced at the
base of the modelled convection zone, in entropy and pressure
equilibrium. The lack of mechanical equilibrium thus causes the tube to
be initially buoyant and rise through the convection zone up the the
top of the computational domain. In this work, the interactions between
the magnetic tube-like structure and convective motions were
investigated for different initial field strength and these results
were compared to isentropic cases (where the convective instability
is not triggered). We do not go into the details of the simulations
and the results. We just wish to summarise some of the findings of these 3D calculations
which will be useful for the present work.

\subsection{The rise velocity}

As shown in \cite{Jouve09}, while it rises, the flux tube creates its own local
velocity field which may strongly disturb the background velocity
field, especially when the initial magnetic field intensity is strong
compared to that of the equipartition field $B_{eq}$ (defined as the field
which energy is equal to the kinetic energy of the strongest downdrafts). This explains why the
tube is more or less influenced by the convective motions as it
evolves in the CZ. Indeed, if the initial magnetic field on the axis
of the rope is $6\times10^5 \, \rm G$ (corresponding to 10 $B_{eq}$), the
tube creates a velocity through the action of the Lorentz force which
dominates against the background velocity. Since a strong upflow is thus created near the tube axis,
the rising mechanism is very efficient and the tube reaches the top of
the CZ in only 3 to 4 days. In the weaker case (where the initial
field strength is $3\times10^5 \, \rm G$, corresponding to 5 $B_{eq}$)  , the velocity field
created by the magnetic structure is comparable to the background
velocity field and the latter is thus able to influence the behaviour
of the flux rope as it rises, the rise time is in this case of about
12 days.

\begin{figure}[h!]
	\centering
	\includegraphics[width=8.cm]{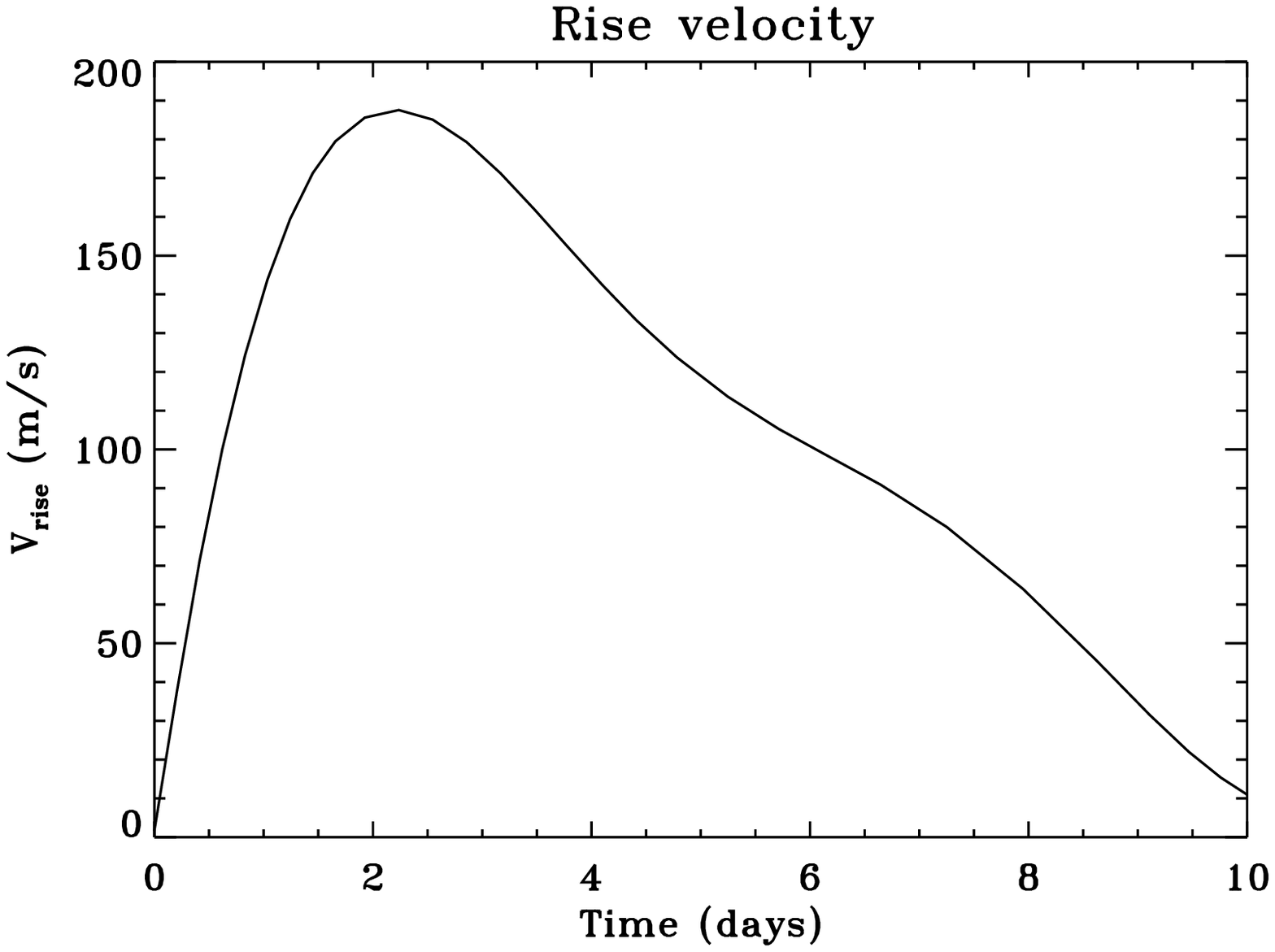}
	\includegraphics[width=8.cm]{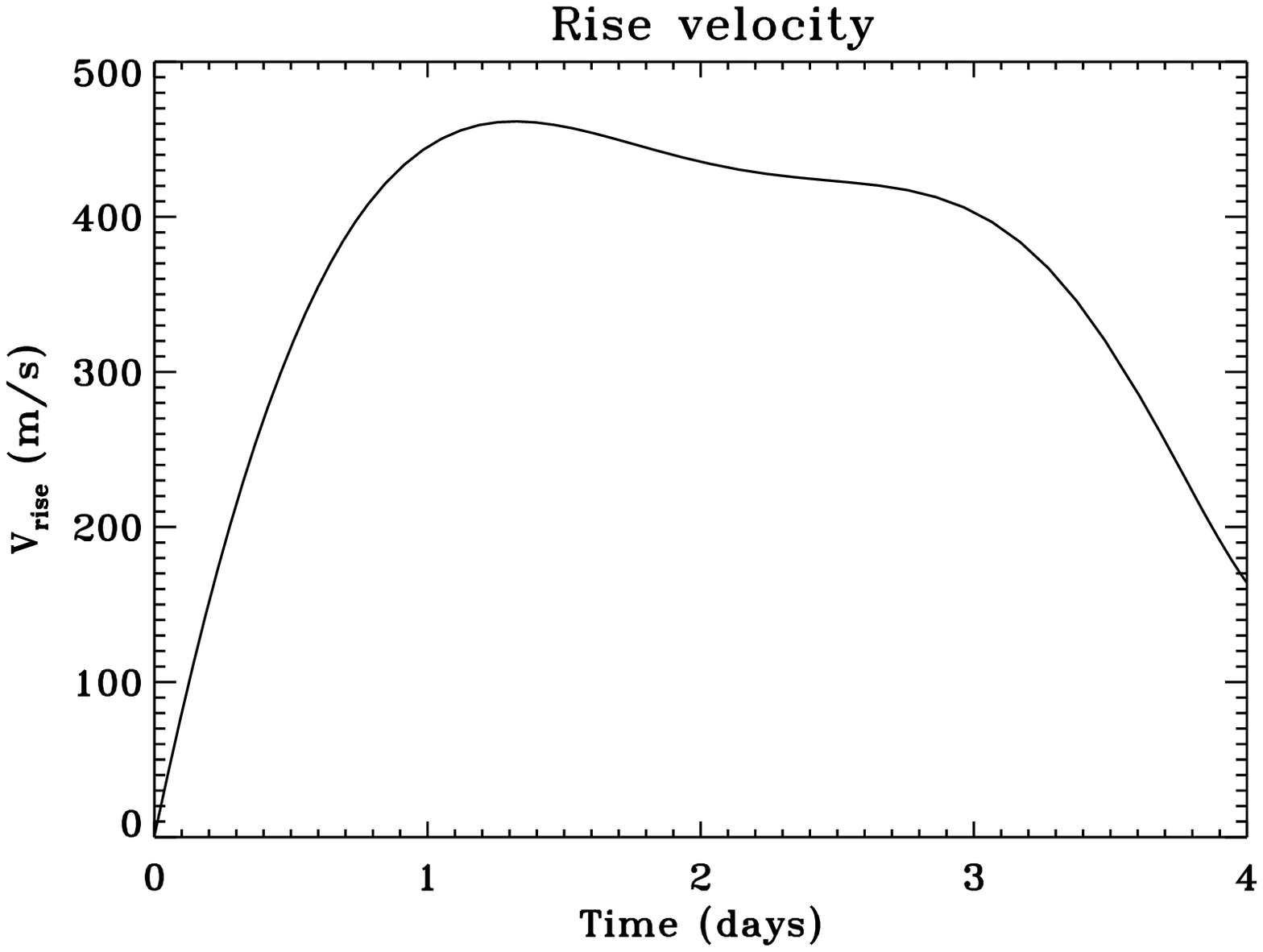}
	\caption{Rise velocity of tubes
	introduced at 5 $B_{eq}$ (upper panel) and 10 $B_{eq}$ (lower
	panel) measured as a function of time during the evolution of
	the tubes from the base of the CZ up to about $0.9 R_{\odot}$.}
	\label{figure_risev}
\end{figure}

Figure \ref{figure_risev} shows more precisely the evolution of the
rise velocity of tubes introduced at the two different field strengths
discussed above. On the
upper panel, the rise velocity for the 5 $B_{eq}$-field is shown and
the lower panel shows the evolution for the stronger field (10
$B_{eq}$). We first note the different time scales for the tubes to
reach the surface, and this is the particular feature which will be
introduced in 2D mean-field models. This figure also shows that the
acceleration phase of these two tubes are similar, but after the
maximal velocity has been reached, the behaviour of the two tubes
differs. In particular, the weak tube gets strongly influenced by the
convective motions and magnetic diffusion. As a consequence, its rise
velocity is significantly reduced as it proceeds through the convection
zone. On the contrary, after the strong tube has reached its maximal
velocity, it keeps making its way to the surface at the same speed and
the strong decrease shown on the figure is eventually strongly linked
to the upper boundary condition (the boundary conditions for the
velocity are stress-free and impenetrable in this case).

\section{Reintroduction in 2D models}

The idea of this work is to design a new 2D  mean-field flux transport
Babcock-Leighton model which takes into account the findings of 3D
calculations concerning the rise of strong toroidal structures from
the base of the convection zone up to the surface. In this section, we
thus present the usual mean-field equations and the basic ingredients
used in the model, with a particular focus on the new Babcock-Leighton
surface source term.

\subsection{The mean field Babcock-Leighton model}

To model the solar global dynamo, we use the hydromagnetic induction equation, governing the evolution of the magnetic field ${\bf B}$ in response to advection by a flow field ${\bf V}$ and resistive dissipation.

\begin{equation}
\frac{\partial {\bf B}}{\partial t}=\nabla\times ({\bf V} \times{\bf B})-\nabla\times(\eta\nabla\times{\bf B}) 
\end{equation}
where $\eta$ is the magnetic diffusivity.

As we are working in the framework of mean-field theory, we express both magnetic and velocity fields as a sum 
of large-scale (that will correspond to mean field) and small-scale (associated with fluid turbulence) contributions. 
Averaging over some suitably chosen intermediate scale makes it possible to write two distinct induction equations for 
the mean and the fluctuating parts of the magnetic field. A closure
relation is then used to express the electromotive force 
in terms of mean magnetic field, leading to a simplified mean-field equation. In this work we will replace the emf by a surface
Babcock-Leighton term (Babcock 1961; Leighton 1969; Wang et al. 1991;
Dikpati \& Charbonneau 1999; Jouve \& Brun 2007a) as described in details below.

% The mean-field equation reads

%\begin{eqnarray}
%\frac{\partial {\bf \langle B\rangle}}{\partial t}&=&\nabla\times ({\bf \langle V\rangle} \times{\bf \langle B\rangle})
%+\nabla\times\langle{\bf v'}\times {\bf b'}\rangle \nonumber \\
%&-&\nabla\times(\eta\nabla\times{\bf \langle B\rangle})
%\end{eqnarray}

%\noindent where ${\bf \langle B\rangle}$ and ${\bf \langle V\rangle}$ refer to the mean parts of the magnetic and velocity fields and ${\bf v'}$ and ${\bf b'}$ 
%to the fluctuating components. A closure relation is then used to express the electromotive force $\langle{\bf v'}\times {\bf b'}\rangle$ 
%in terms of mean magnetic field, leading to a simplified mean-field equation. In this work we will replace the emf by a surface
%Babcock-Leighton term (Babcock 1961; Leighton 1969; Wang et al. 1991;
%Dikpati \& Charbonneau 1999; Jouve \& Brun 2007a) as described in details below.

%Working in spherical coordinates and under the assumption of axisymmetry, we write the total mean magnetic field {\bf B} 
%and the velocity field {\bf V} as
%\begin{equation}
%{{\bf B}}(r,\theta,t)=\nabla\times (A_{\phi}(r,\theta,t) \hat {\bf e}_{\phi})+B_{\phi}(r,\theta,t) \hat {\bf e}_{\phi}
%\end{equation}
%\begin{equation}
%{{\bf V}}(r,\theta)={\bf v_{p}}(r,\theta) + r\sin\theta \, \Omega(r,\theta) \hat {\bf e}_{\phi}
%\end{equation}

%Note that our velocity field is time-independent since we will not assume any fluctuations in time of the differential 
%rotation $ \Omega$ or of the meridional circulation ${\bf v_{p}}$.

We work here in spherical
  geometry and under the assumption of axisymmetry. 
Reintroducing a poloidal/toroidal decomposition of the magnetic and
  velocity fields in the mean induction equation, we get two coupled 
partial differential equations, one involving the poloidal potential $A_{\phi}$ and the other concerning the toroidal field $B_{\phi}$. 

\begin{eqnarray}
\label{eqA2}
\frac{\partial {A_{\phi}}}{\partial t}&=&\frac{\eta}{\eta_{t}} (\nabla^{2}-\frac{1}{\varpi^{2}})A_{\phi}-
R_{e}\frac{\bf{v}_{p}}{\varpi}\cdot\nabla(\varpi A_{\phi}) \nonumber
\\
&+&C_{s}S(r,\theta,B_{\phi})
\end{eqnarray}

\begin{eqnarray}
\label{eqB2}
\frac{\partial {B_{\phi}}}{\partial t}&=&\frac{\eta}{\eta_{t}} (\nabla^{2}-\frac{1}{\varpi^{2}})B_{\phi}
+\frac{1}{\varpi}\frac{\partial(\varpi B_{\phi})}{\partial r}\frac{\partial (\eta/\eta_{t})}{\partial r} \nonumber  \\
&-&R_{e}\varpi {\bf v}_{p}\cdot\nabla(\frac{B_{\phi}}{\varpi})-R_{e}B_{\phi}\nabla\cdot{\bf v}_{p} \nonumber  \\ 
&+&C_{\Omega}\varpi(\nabla\times(\varpi A_{\phi}{\bf \hat{e}}_{\phi}))\cdot\nabla\Omega
\end{eqnarray}

\noindent where $\varpi=r\sin\theta$, $\eta_{t}$ is the turbulent magnetic diffusivity (diffusivity in the convective zone), ${\bf v}_{p}$ 
the flow in the meridional plane (i.e. the meridional circulation), $\Omega$ the differential rotation, $S(r,\theta,B_{\phi})$ 
the Babcock-Leighton source term for poloidal field.
In order to write these equations in a dimensionless form, we choose as length scale the solar radius ($R_{\odot}$) and as time 
scale the diffusion time ($R_{\odot}^2/\eta_{t}$) based on the envelope diffusivity ($\eta_{t}$).
This leads to the appearance of three control parameters $C_{\Omega}=\Omega_{0}R_{\odot}^2/\eta_{t}$, $C_{s}=s_{0}R_{\odot}/\eta_{t}$ 
and $R_{e}=v_{0}R_{\odot}/\eta_{t}$ where $\Omega_{0}, s_{0}, v_{0}$ are respectively the rotation rate and the typical amplitude of  
the surface source term and of the meridional flow.

Equations $\ref{eqA2}$ and $\ref{eqB2}$ are solved in an annular meridional cut with the colatitude $\theta$ $\in [0,\pi]$ and the 
radius (in dimensionless units) $r \in [0.6,1]$ i.e from slightly below the tachocline ($r=0.7$) up to the solar surface, using the STELEM code.
This code has been thoroughly tested and validated thanks to an international mean field dynamo benchmark involving 8 different codes (\cite{Jouve08}). 
At $\theta=0$ and $\theta=\pi$ boundaries, both $A_{\phi}$ and $B_{\phi}$ are set to 0. Both $A_{\phi}$ and $B_{\phi}$ are set 
to $0$ at $r=0.6$. At the upper boundary,  we smoothly match our solution to an external potential field, i.e. we have vacuum 
for $r \geq 1$.
As initial conditions we are setting a confined dipolar field configuration, i.e the poloidal field is set to $\sin\theta / r^{2}$ in 
the convective zone and to $0$ below the tachocline whereas the toroidal field is set to $0$ everywhere. \\

\subsection{The standard physical ingredients and the modified surface
source term}

We now need to prescribe the amplitude and profile of the various ingredients acting in
equations $\ref{eqA2}$ and $\ref{eqB2}$, namely the rotation,
the diffusivity, the meridional flow and the Babcock-Leighton source term.

The rotation profile captures some realistic aspects of the Sun's angular velocity, deduced from helioseismic inversions (Thompson et al. 2003), assuming 
a solid rotation below $0.66$ and a differential rotation above the interface. 

\begin{eqnarray}
{\Omega(r,\theta)}&=&\Omega_{c}+\frac{1}{2}[1+\rm erf(2\frac{r-r_{c}}{d_{1}})] \nonumber \\
& &(\Omega_{Eq}+a_{2}\cos^2\theta+a_{4}\cos^4\theta-\Omega_{c})
\end{eqnarray}

\noindent with $\Omega_{Eq}=1$,  $\Omega_{c}= 0.93944$,  $r_{c}=0.7$,
$d_{1}=0.05$, $a_{2}=-0.136076$ and $a_{4}=-0.145713$ and where $\rm erf$
is the error function.
With this profile, the radial shear is maximal at the tachocline.

%\begin{figure}[!h]
%  \centering
%\includegraphics[width=8cm]{./omega_std.ps}
%\includegraphics[width=8cm]{./vit_1cel_bench.ps}
%\caption{Differential rotation and meridional circulation used in the standard models. Plain (dashed) lines indicate counterclockwise (clockwise) circulation.} 
%\label{domega}
%\end{figure}

We assume that the diffusivity in the envelope $\eta$ is dominated by its turbulent contribution whereas in the stable interior 
$\eta_{\rm c} \ll \eta_{\rm t}$. We smoothly match the two different constant values with an error function which enables us to 
quickly and continuously transit from $\eta_{\rm c}$ to $\eta_{\rm t}$ i.e.

\begin{equation}
\eta(r)=\eta_{\rm c}+\frac{1}{2}\left(\eta_{\rm t}-\eta_{\rm c}\right)\left[1+{\rm erf}\left(\frac{r-r_{\rm c}}{d}\right)\right],
\label{eqeta}
\end{equation}

\noindent with  ${\eta_{\rm c}}=10^9 \,\rm cm^2\rm s^{-1}$ and $d=0.03$.

In Babcock-Leighton flux-transport dynamo models, meridional circulation is used to link the two sources of the magnetic field namely 
the base of the CZ and the solar surface.
For all the models presented in this paper we use a large single cell per hemisphere, directed poleward at the surface, vanishing at the bottom boundary 
$r=0.6$ and thus penetrating a little below the tachocline.
We take a stream function

\begin{equation}
\psi(r,\theta)=-\frac{2}{\pi}\frac{(r-r_{b})^2}{(1-r_{b})}\sin\left(\pi\frac{r-r_{b}}{1-r_{b}}\right)\cos\theta\sin\theta,
\label{eqpsi}
\end{equation}

\noindent which gives, through the relation  ${\bf u_{\rm p}}=\nabla \times(\psi \hat {\bf e}_{\phi})$, the following components of the meridional flow

\begin{eqnarray}
u_{r}&=&-\frac{2(1-r_{\rm b})}{\pi r}\frac{(r-r_{\rm b})^2}{(1-r_{\rm b})^2} \sin\left(\pi\frac{r-r_{\rm b}}{1-r_{\rm b}}\right)(3\cos^2\theta-1), 
\end{eqnarray}
\begin{eqnarray}
u_{\theta}&=&\Bigg[\frac{3r-r_{\rm b}}{1-r_{\rm b}} \sin\left(\pi\frac{r-r_{\rm b}}{1-r_{\rm b}}\right)+\frac{r\pi}{1-r_{\rm b}}\frac{(r-r_{\rm b})}{(1-r_{\rm b})} \cos\left(\pi\frac{r-r_{\rm b}}{1-r_{\rm b}}\right)\Bigg] \nonumber \\
	  & \times &\frac{2(1-r_{\rm b})}{\pi r}\frac{(r-r_{\rm b})}{(1-r_{\rm b})}\cos\theta\sin\theta, 
\end{eqnarray}
\noindent with $r_{\rm b}=0.6$.

%\subsection{The modified source term for poloidal field}

%\bigskip

In Babcock-Leighton dynamo models, the poloidal field owes its origin to the tilt of magnetic loops emerging at 
the solar surface. Since these emerging loops are thought to rise from
the base of the convection zone through magnetic buoyancy, we see that
we can directly relate the way we model the Babcock-Leighton (BL) source term and the
results of 3D calculations shown in Sect \ref{sect_3D}.

%Thus, the source has to be confined to a thin layer just below the surface and since the process is fundamentally 
%non-local, the source term depends on the variation of $B_{\phi}$ at
%the base of the convection zone. 

In the standard model, the source term is confined in a thin layer at
the surface and is made to be antisymmetric with respect to the
equator, due to the sign of the Coriolis force which changes from one
hemisphere to the other. These features are retained in our modified
version.

However, the standard term is proportional to the toroidal field at
the base of the convection zone at the same time, implying an
instantaneous rise of the flux tubes from the base to the surface
where they create tilted active regions. The 3D calculations showed
that the rise velocity and thus the rise time depend on the
field strength at the base of the CZ, we thus introduce in our
modified version of the source term, a magnetic field-dependent time
delay in the toroidal field at the base of the convection zone. We
thus take into account the time delay between the formation of strong
toroidal structures at the base of the convection zone and the surface
regeneration of poloidal field.

%The second modification concerns the latitude of emergence of active
%regions responsible for the regeneration of poloidal field. In
%\ref{sect_3D}, we reminded the reader that the latitude of emergence
%depends on the toroidal field strength at the base of the convection
%zone. In a crude way, we can model this by introducing a threshold for
%the field strength above which the flux tubes will rise radially and
%below which they will rise vertically. This threshold is taken to be
%equal to $10^5 \, \rm G$, in agreement with 3D calculations. 

The modified expression of the source term is thus

\begin{eqnarray}
S(r,\theta,B_{\phi})&=& \frac{1}{2}\Bigg[1+\rm erf(\frac{r-r_{2}}{d_{2}})\Bigg]\Bigg[ 1-erf(\frac{r-1}{d_{2}})\Bigg] \nonumber \\ 
& \times & \Bigg[1+({\frac{B_{\phi}(r_{c},\theta,t-\tau_B)}{B_{0}}})^{2}\Bigg]^{-1}
\nonumber \\
& \times &\cos\theta \sin\theta \,\, B_{\phi}(r_{c},\theta,t-\tau_B) 
\label{eq_s}
\end{eqnarray}

where $r_{2}=0.95$, $d_{2}=0.01$, $B_{0}=10^4 \, \rm G$, with the time
delay $\tau_B$ proportional to the inverse of the magnetic energy at
the base of the convection zone,
namely 
$$
\tau_B(\theta,t) = \tau_0/B_{\phi}(r_c,\theta,t)^2
$$

 In our
3D simulations, the approximate rise time for a $6 \times 10^5 \, \rm
G$ field is indeed about 4 times that of a $3 \times 10^5 \, \rm G$
field. This takes into account the more significant effects of
convective downdrafts and Ohmic diffusion in the ``weak B'' case. Note
that the time delay is then dependent both on space and time.

%To take into account the latitudinal drift of our rising toroidal
%structures, we introduce 2 contributions to our surface term at
%$\theta$, one which comes from strong tubes (i.e. more than $10^4\,
%\rm G$) emerging from the same
%colatitude $\theta$ and one which comes from weak tubes (i.e. less than $10^4 \,
%\rm G$) emerging from the colatitude $\psi$ such that
%$\sin(\psi)=R_\odot/r_c \sin(\theta)$. Of course, this contribution will
%exist only for colatitude less than $\arcsin(r_c/R_\odot)$ and the
%points at higher colatitudes (lower latitudes) will only get contributions from strong
%radially rising tubes. Note that in this case, the distribution of
%toroidal field at the base of the convection zone will not be
%representative of the location of the sunspots appearing at the
%surface. Instead, we have now a mapping between the toroidal field at
%the base of the CZ and the butterfly diagram, which represents the
%distribution of sunspots at the solar surface. 

Finally, we take into account the fact that very strong toroidal
structures (more than $10^5 \, \rm G$) are not influenced by the
Coriolis force enough to gain a significant tilt as they reach the
surface. To do so, we introduce a quenching term in the surface source
which will make the regeneration of the poloidal field less effective
when the toroidal field at the base of the CZ is strong enough.
Inversely, when the flux tubes are too weak, they will not be able to
reach the surface at all, and will not take part into the regeneration
term for the poloidal field. %We thus introduce a lower threshold  
We thus prevent the weakest flux tubes (or equivalently the most
delayed ones) to take part in the regeneration of poloidal fields at
the surface. To do so, the source term is set to zero when the delays
are above a certain threshold value corresponding to a rise time of
approximately half a solar cycle.

\section{Results}
\label{sect_2D}
\subsection{The standard solution}

We first quickly present the solutions of our dynamo equations for the
standard case which has been extensively discussed and used in the
community to model the solar cycle (e.g. Dikpati $\&$ Charbonneau, 1999). 

\begin{figure}[h!]
	\centering
	\includegraphics[width=9cm]{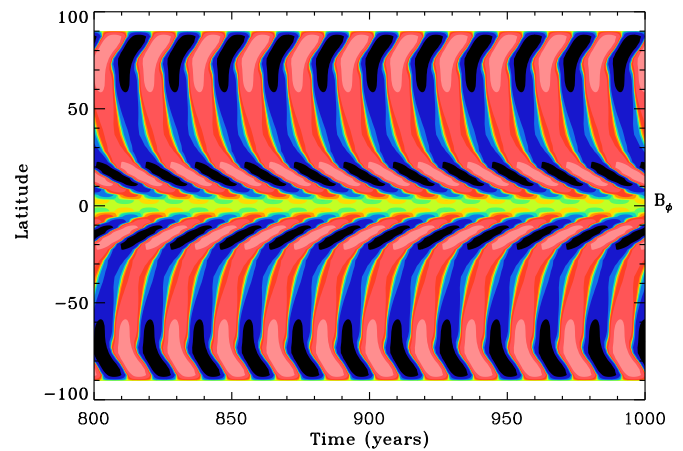}
	\includegraphics[width=9cm]{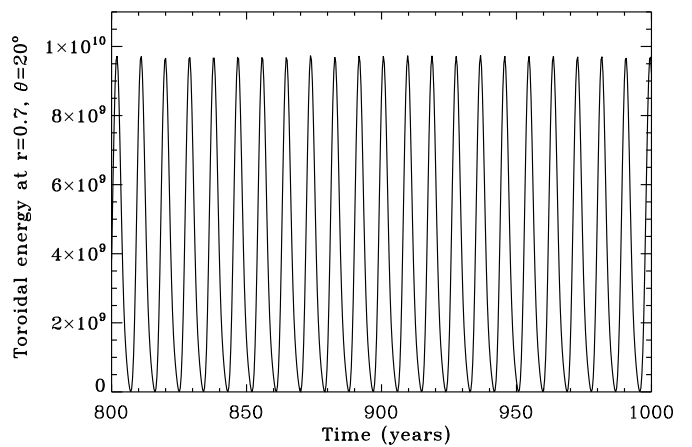}
	\caption{Time-latitude cut of the toroidal field at the base
	of the convection zone (representing the butterfly diagram)
	and temporal evolution of the toroidal field energy $(B_\phi^2)$ at a particular
	position in space (at the base of the convection zone and at
20\degr in latitude in the Northern hemisphere). Time is in years and
the magnetic field in Gauss. Red (blue) colours indicate positive
(negative) field values, ranging from $-2 \times 10^5 \, \rm G$ to
  $2 \times 10^5 \, \rm G$.}
	\label{figure_std}
\end{figure}

Figure \ref{figure_std} shows the typical solution of a
Babcock-Leighton standard dynamo model with a realistic choice of
parameters. This is the same case as the reference case of \cite{Jouve07a}. In particular, the maximum latitudinal velocity at the
base of the convection zone (which is the significant velocity for the
advection of strong toroidal fields close to the tachocline) is set to
about $1\,\rm m.s^{-1}$.

The butterfly diagram resulting from this particular model is in good
agreement with the observed migration of sunspots at the surface. In
particular, a strong equatorward branch stretching from the
mid-latitudes to the equator is clearly visible on the upper panel of
Figure \ref{figure_std}. A strong poleward branch also appears in
  the evolution of the toroidal field at the interface, which is quite
  typical of Babcock-Leighton models. Indeed, it is the result of the
  combined action of low diffusivity at the base of the convection
  zone and strong radial shear linked to the differential rotation
  profile. It has been shown that both higher diffusivities
  (\cite{Dikpati99}) and penetration of the meridional flow below the
  core-envelope interface (\cite{Nandy01}) could prevent
  such a branch to appear. However, since our goal
  is more to study the effects of time delays in BL models than to
  build a calibrated solar-like solution, we adopt this solution as
  our reference case. 

Nevertheless, it has here to be pointed out
that making the link between the evolution of toroidal field at the
base of the convection zone and the sunspots location at the solar
surface during a magnetic cycle is not straightforward. Indeed, to be
able to make this link, we have to
assume that the rise of toroidal structures from the base of the
convection zone to the surface is radial. However, as \cite{Choudhuri87} first demonstrated using the
thin flux tube approximation and as \cite{Fan08} confirmed with 3D MHD simulations,
initially weak magnetic flux tubes are deviated from the radial
trajectory and tend to follow a path which is parallel to the rotation
axis while they make their way to the solar surface. In the
simulations presented here, we do not take into account this drift in
latitude of weak structures but intend to do so in a future work. We
thus assume the evolution of toroidal field at the base of the
convection zone to be a good proxy for the sunspot migration at the
surface but keep in mind that the butterfly diagram may
take a different form if the latitudinal drift is taken into account.

The lower panel of Figure \ref{figure_std} shows the time evolution of
the toroidal field energy at the base of the convection zone and at
the latitude of $20\degr$, inside the activity belt. We clearly note that for the choice of
parameters we made, the solution is harmonic in time, no modulation of
the cycle is visible. As pointed out in the introduction,
\cite{Charbonneau052} showed that a variation of the strength of the
Babcock-Leighton poloidal source term could lead to a chaotic behaviour
of the solutions through a sequence of bifurcations. In order to focus
on the influence of the time delays on the modulation of the
cycle, we chose
here a value of $C_s$ which lies in the range of values for which no
modulation of the cycle can arise for the standard solution.

\begin{figure}[h!]
	\centering
	\includegraphics[width=9.cm]{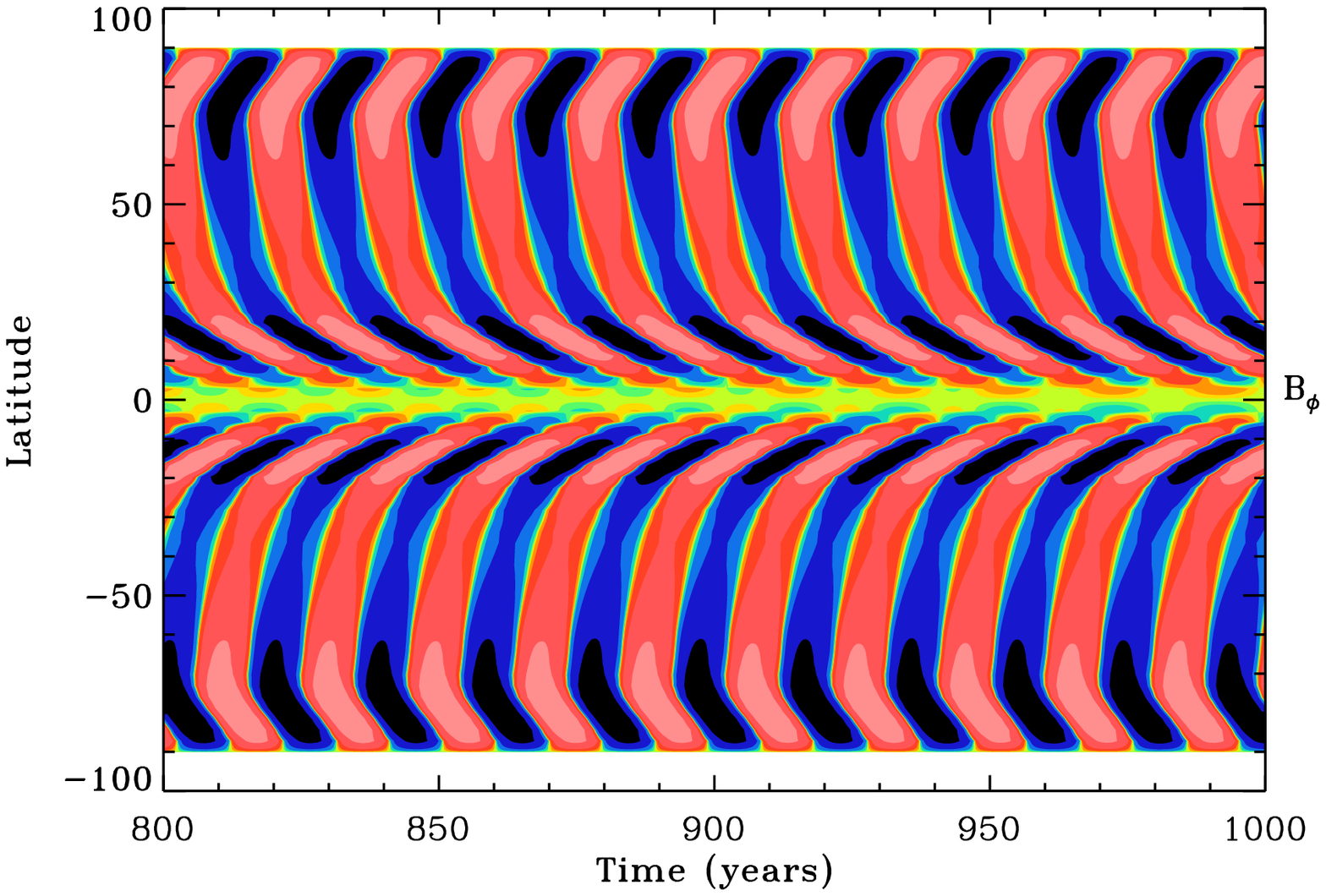}
	\includegraphics[width=9.cm]{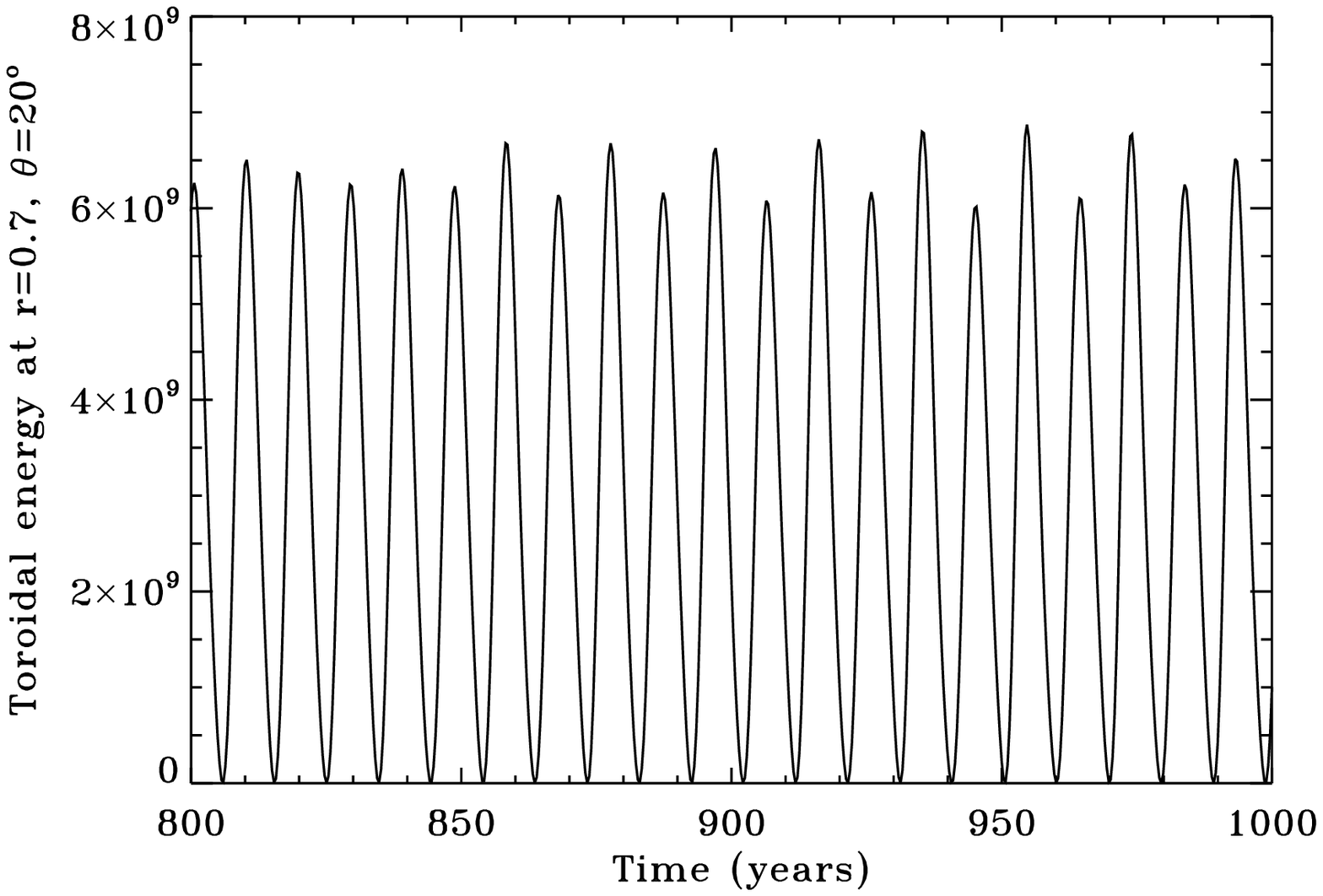}
	\caption{Same as Fig. \ref{figure_std} but with a delay of 14
          days on 1 $\rm kG$ fields.}
	\label{figure_1e3}
\end{figure}

\subsection{Introducing moderate delays}

We now turn to investigate the effect of introducing a magnetic
field-dependent delay in the model, modifying the surface source term
according to Eq. \ref{eq_s}. In particular, we focus on the
modification of the butterfly diagram and the evolution of the
toroidal field energy at a particular point in space.

Figure \ref{figure_1e3} shows the butterfly diagram and toroidal field
energy for a few cycles in the case where a small delay is
introduced. More precisely, a delay of 14 days is imposed on 1 $\rm
kG$ fields at the base of the convection zone. According to the
dependence of the delay on the magnetic field intensity, it means that
10 $\rm kG$ fields will only be delayed by less than 4 hours, which is
extremely small compared to the time scale of the magnetic cycle.
However, even if the effects are hardly visible on the butterfly
diagram (it is almost indistinguishable from the one of
Fig. \ref{figure_std}), a clear modulation of the magnetic energy
appears on the lower panel of Fig. \ref{figure_1e3}. The small
modification of the butterfly diagram is not surprising considering
the fact that the same physical processes still act in our model to
generate and advect the 3 components of the magnetic field. In
particular, the
meridional flow, responsible for the advection of toroidal fields at
the base of the convection zone, still acts to produce the
characteristic equatorward
branch of sunspot migration. What is more striking is the effect on
the cycle amplitude which becomes modulated, while the cycle period
remains approximately constant and identical to the standard case. To understand this
particular feature, we will proceed to the study of a reduced model in
the following section. But first, the effects of higher delays are
analysed.

\subsection{Strong delays leading to highly perturbed solutions}

The time delay is now increased from 14 days on 1 $\rm
kG$ fields to 14 days on 50 $\rm kG$ fields, meaning that 10 $\rm kG$
fields will be delayed by almost a year in this case (compared to a
few hours in the previous calculation). We point out again that the
maximum delay which can be reached for very weak tubes is about 6
years (approximately half a cycle), since a longer time for toroidal
structures to cross the convection zone would be unrealistic. 

\begin{figure}[h!]
	\centering
	\includegraphics[width=9cm]{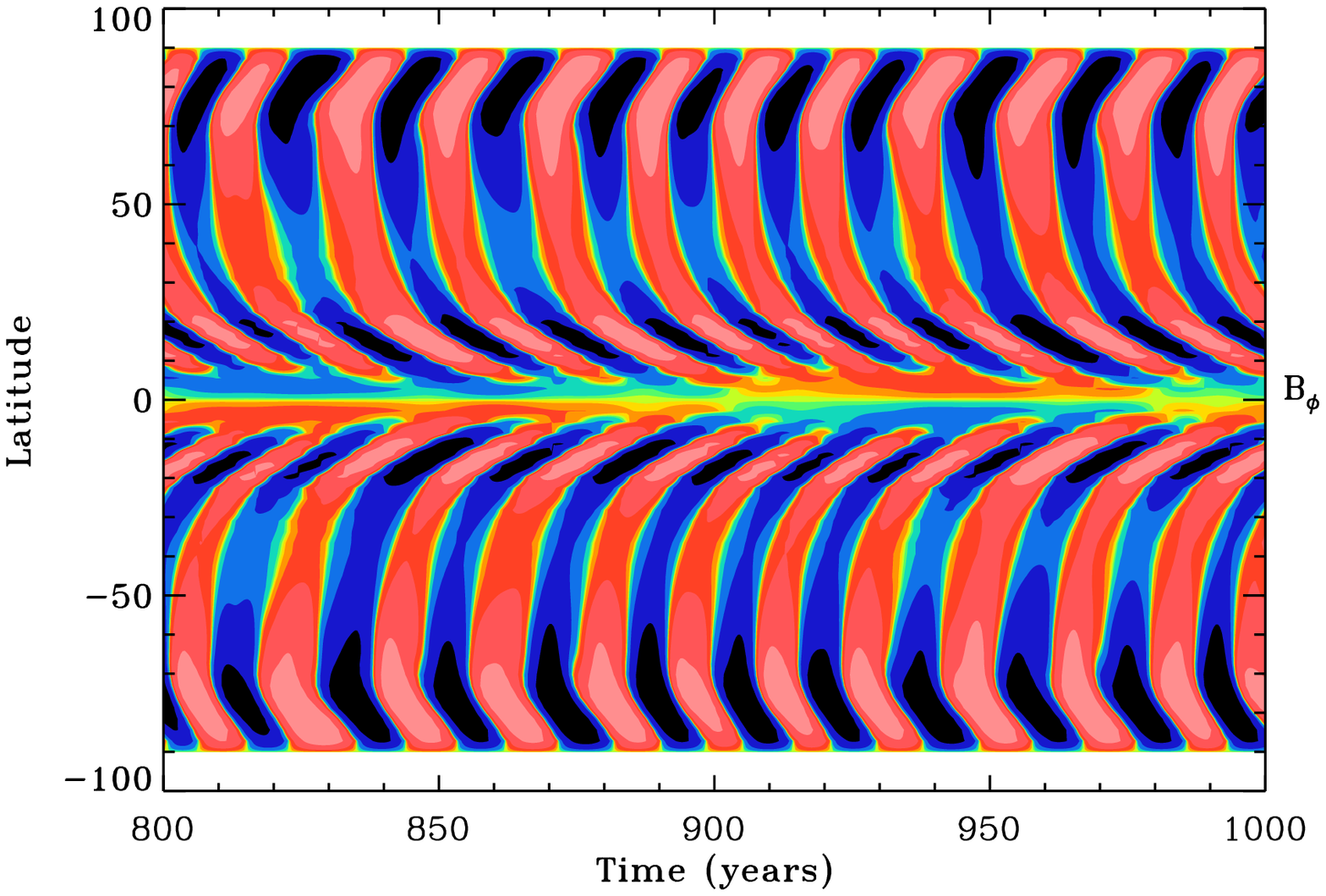}
	\includegraphics[width=9cm]{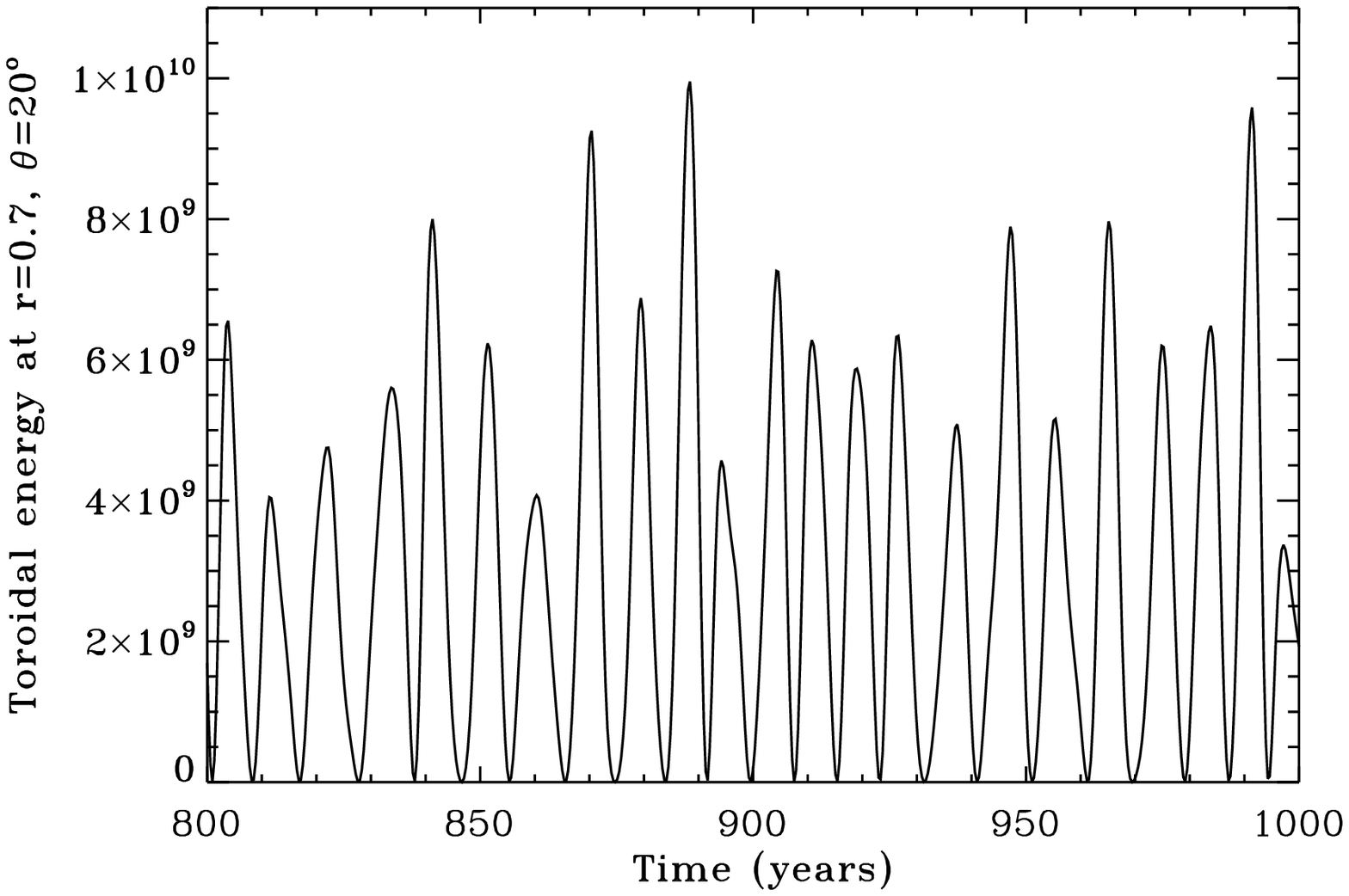}
	\caption{Same as Fig. \ref{figure_std} but with a delay of 14
          days on 50 $\rm kG$ fields.}
	\label{figure_5e4}
\end{figure}

Figure \ref{figure_5e4} shows the butterfly diagram and the evolution
of the magnetic energy for this case.
This case is particularly interesting when we turn to consider the
evolution of the toroidal energy with time. Indeed, again the effects
on the shape of butterfly diagram stay weak, even if some perturbations
are clearly visible. In particular, close to the equator, consecutive
cycles of the same polarity seem to connect, leading the cycle period
at these latitudes to be different from the basic 11-yr cycle still
present at higher latitudes. We also note that the period varies from
one cycle to another but this particular feature is even more visible
on the lower panel of Figure \ref{figure_5e4}.

The major effect of the delays is again to modulate the amplitude of
the cycle to such a point that the weakest cycle on the particular
period shown on Figure \ref{figure_5e4} reaches in terms of peak
energy only 30 $\%$ of the peak energy of the strongest cycle. These
variations are comparable to the observations of the monthly sunspot
number available from 1750 until today (e.g. \cite{Hoyt98}). Indeed, over this period of
time, the weakest cycle (cycle 5, which reached its peak in 1804)
has been shown to possess a sunspot number of about 50 whereas the
strongest cycle (cycle 19, which reached its maximum value in 1955)
reached a sunspot number of more than 200. These strong variations are
thus reproduced in this model with rather strong delays, but still
small compared to the cycle period.

Finally, we should stress the variety of morphologies exhibited by the
different cycles. In particular, some cycles show a clear asymmetry
between their rising and declining phases (for example the cycle peaking at
t=895). Interestingly, we also note that some
cycles seem to possess almost two peaks (as was
observed on cycle 23) or at least show a change of slope in the
declining phase, which makes them quite different from the standard
more ``classical'' cycle.

%\subsection{Memory of the system}

%\section{Interpretation and discussion}

To better understand the results of such a modulation in the amplitude
and possibly on the period
of the modelled solar cycle, we need to get back to a reduced model
and analyse its properties from a dynamical systems point of view.

\section{Understanding the modulation: reduction to a 6th order
  system}
\label{sect_6th}

\subsection{Formulation and justification of the reduced model}
\label{6thorder}

The reduced model we decide to use does not explicitly contains
time-delays as in the 2D model. It rather introduces another variable
$Q$ which will be delayed by time $\tau$ with respect to the toroidal field
$B$. Moreover, we now work in only one dimension in space.
The set of equations we thus want to solve here is the following:

\begin{equation}
\frac{\partial A}{\partial t}+ v \frac{\partial A}{\partial x}= \frac{S Q}{1+\lambda \vert Q \vert^2}+\eta
\frac{\partial^2 A}{\partial x^2}
\end{equation}

\begin{equation}
\frac{\partial B}{\partial t}+ v \frac{\partial B}{\partial x}=\Omega \frac{\partial A}{\partial
  x}+\eta \frac{\partial^2 B}{\partial x^2}
\end{equation}

\begin{equation}
\frac{\partial Q}{\partial t}= \frac{1}{\tau}(B-Q)
\end{equation}

\noindent This model is significantly simplified since the physical
ingredients (the rotation rate, the magnetic diffusivity, the
meridional flow and the
source term) are taken to be constants. $\tau$ is still dependent on the inverse of the 
toroidal field $B$, leading to the following expression: 

$$
\tau=\frac{\tau_0}{1+\vert B \vert ^2}
$$

\noindent We will nevertheless show that such a simple model with a varying time
delay, even small, is able to reproduce a modulation in the cyclic
behaviour of the solutions. 

In order to simplify even more the previous set of equations, we take
the variables $A$, $B$ and $Q$ to be single Fourier modes in $x$,
with wavenumber k, such that:
$$
{A}(x,t)=A_t(t) \exp(i k x) 
$$
$$
{B}(x,t)=B_t(t) \exp(i k x)
$$
$$
{Q}(x,t)=Q_t(t) \exp(i k x)
$$

\noindent we
then get the following set of complex ODEs:

\begin{equation}
\frac{d A_t}{d t}+ i k v A_t= \frac{S Q_t}{1+\lambda \vert Q_t
  \vert^2}-\eta k^2 A_t
\label{eqA}
\end{equation}

\begin{equation}
\frac{d B_t}{d t}+ i k v B_t=i k \Omega A_t-\eta k^2 B_t
\label{eqB}
\end{equation}

\begin{equation}
\frac{d Q_t}{d t}= \frac{1}{\tau}(B_t-Q_t)
\label{eqQ}
\end{equation}

\noindent with $\tau=\tau_0/(1+\vert B_t \vert ^2)$.
All the analysis of the previous set of ODEs will be made considering
$k=1$ to be the wavenumber of interest in our model. This system
possesses 6 degrees of freedom represented by the real and imaginary
parts of each variables $A_t$, $B_t$ and $Q_t$.

%\bigskip

This choice of model was motivated by the fact that we want to study
this set of ODEs as a dynamical system. Having explicit delays in the equations
makes it much more difficult to identify the sequence of bifurcations
which could occur between the harmonic solution of
Fig. \ref{figure_std} and the aperiodic solution of
Fig. \ref{figure_5e4}. Nevertheless, the new variable $Q_t$ introduced
in our simplified model and which now
represents the delayed toroidal field behaves exactly as we want it
to. To illustrate this, we solve the set of equations numerically for
the following choice of parameters: $\Omega=100$, $S=100$,
$\eta=10^{-3}$, $v=10^{-2}$, $\lambda=10^{-4}$ and $\tau_0=2\times
10^{-2} B_{1}^2$ where $B_{1}=3\times 10^8$ is a normalization factor. From now on, the values of $\tau_0$
will be expressed in terms of multiples of $B_{1}^2$. \emph{We note
  here that the actual time delays should not be directly related to the values of
$\tau_0$, beacuse of this normalization factor.}

These parameters are difficult to relate to the ones used in the full 2D model
but they were chosen so that:
\begin{itemize}
\item dynamo action occurs (the dynamo number
is high enough),
\item the standard solution is harmonic in time (the dynamo
number is low enough so it does not lead to any modulation of the
cycle),
\item we lie in the advection-dominated regime rather than in the
diffusion-dominated one (the advection time $1/k v$ is shorter than the
diffusive time $1/k^2 \eta$, leading to $v > \eta$ in our case where $k=1$),
\item the time delay is always smaller than half a cycle period
  (i.e. we impose a maximum value for the time delay, like we did in
  the 2D model where the source was set to zero when the delay was too long).

\end{itemize}

%These constraints on the parameters were thus quite significant and  

\begin{figure}[h!]
	\centering
	\includegraphics[width=8.cm]{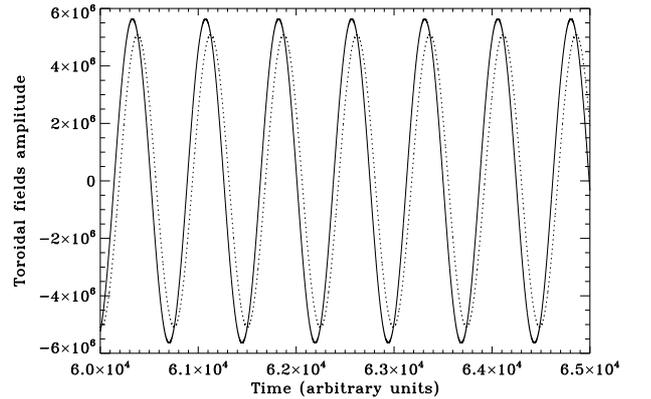}
	\caption{Temporal evolution of the toroidal field (plain line)
        and delayed toroidal field (dotted line) for a case where the
        delay is small enough to keep the harmonic solution stable.}
	\label{figure_bq}
\end{figure}

Figure \ref{figure_bq} shows the evolution over a
few cycles of the toroidal field $\Re(B_t)$ and the delayed
toroidal field $\Re(Q_t)$. We can notice two major particularities:
first, we clearly note the time shift between the two fields,
representing the time delay imposed by the value of $\tau_0$. In this
case, the solution is harmonic and hence $ \vert B_t \vert ^2 $ is
constant and thus the time delay is also constant. Consequently, the time
shift between the two fields is always the same along the different
cycles which are represented on the figure.

The second point which has to be noticed is the significantly lower
amplitude of $\Re(Q_t)$ compared to the original toroidal field
$\Re(B_t)$. This amplitude difference is due to the fact that the
evolution equation for $Q_t$ is similar to a diffusion equation. This
property models the ohmic diffusion acting on the toroidal
flux tubes during their rise through the convection zone. Indeed, we
can reasonably expect the flux tubes reaching the surface to have a lower amplitude
than at the base of the convection zone. Moreover, the decrease in
strength will be higher when the initial magnetic field intensity in
the flux tubes is lower. Indeed, when the flux tubes are weaker, their
rise time is longer and becomes comparable to the diffusive time
associated to the magnetic structure, which is equal to $a^2/\eta$ when
the rising flux tube has a circular section of radius $a$. In our
reduced model, this effect is also taken into account since the
the time delay for weaker tubes will be higher and thus the diffusion
operator will be more efficient in this case.

With this choice of model and parameters, we are thus close to the
dynamo solution
calculated with the STELEM code presented in the previous section, so
that we can reasonably relate
the results of the reduced model to the full 2D calculations. The
major interest of this type of models is that we are now reduced to a
6th order system which can be analysed both analytically and
numerically.

\subsection{Destabilization of the harmonic solution and influence of parameters}

We now proceed to the analytic study of the characteristics of the harmonic
solution when the delay is increased. In particular, we
investigate the influence of the model parameters on the amplitude and
frequency of the harmonic solution. The details of this analysis are
presented in appendix \ref{freq}.

 As expected, the dynamo number $\Omega
S/\eta$ controls the amplitude of our solution while the meridional
flow speed sets the cycle period. This behaviour is characteristic of
mean-field flux transport dynamo model and we are thus not going into
too many details. The effects of an increasing delay on both the
frequency and the amplitude of the solution is on the contrary of some
interest for our particular study.

\begin{figure}[h!]
	\centering
	\includegraphics[width=4.45cm]{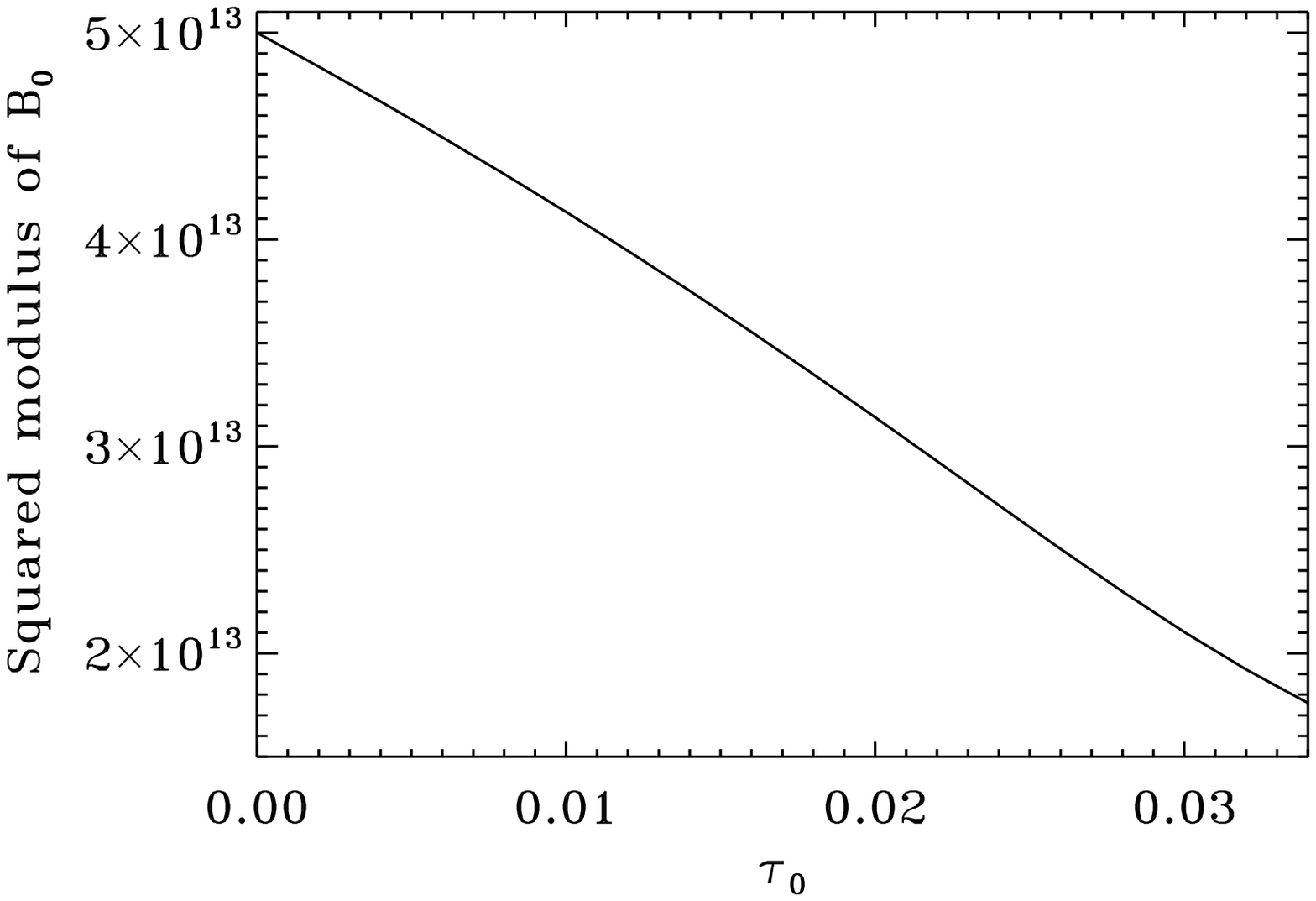}
	\includegraphics[width=4.45cm]{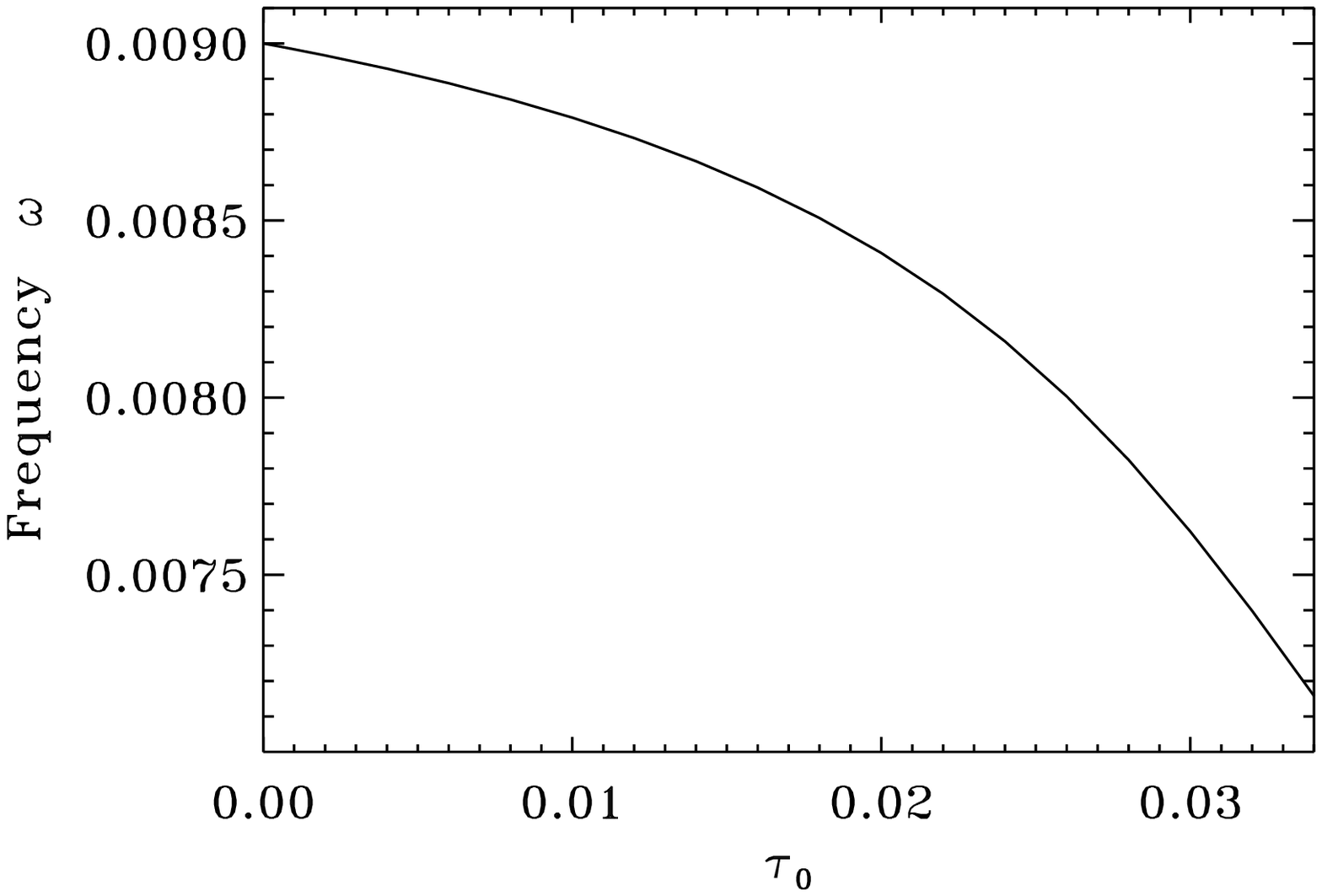}
	\caption{Variation of the amplitude and the frequency of the
          toroidal field $B_t$ with the delay, for a particular
          case. The choice of parameters is the same as in the
          previous section, namely: $\Omega=100$, $S=100$,
$\eta=10^{-3}$, $v=10^{-2}$, $\lambda=10^{-4}$.}
	\label{figure_amp}
\end{figure}

Figure \ref{figure_amp} shows the variation of the amplitude and
frequency of the harmonic solution when the delay is increased when
all the other parameters are fixed to the values quoted in the
previous section. For higher values of the delay, the solution
becomes unstable, as we will see in the stability analysis below. The
standard ``undelayed'' solution is also included in these plots, for
$\tau=\tau_0=0$. We note that the amplitude of our solution decreases
linearly with the delay which can be explained by the fact that the
amplitude of $\Re(Q_t)$ (which represents the delayed toroidal field
and thus is the one responsible for the regeneration of the poloidal
field $A_t$) is smaller than the amplitude of $B_t$, due to the
diffusive effects included in our equations. Hence, the amount of
poloidal field generated by the reduced toroidal field will be weaker
and consequently, the toroidal field produced through the
$\Omega$-effect acting on the poloidal field will also be reduced.

The frequency of the solution has a different behaviour when $\tau$ is
increased. Indeed, the variation is first close to linear with $\tau$,
meaning that the cycle period is increased linearly with the delay but
then the slope of the curve changes and the cycle is significantly
slowed down by a variation of the delay. This feature can also be
understood by the fact that the regeneration of poloidal field will
take longer and longer when the delay is increased and thus the dynamo
loop will take longer to close.

Another interesting point here is that we can study the effect of the
various parameters on the threshold above which the harmonic solution
becomes unstable. We will not focus on the effect of
the dynamo number on this threshold since it has been studied in
similar conditions before. Instead, we fix the dynamo number and we investigate the evolution of
this threshold when the meridional flow speed is increased. The
detailed stability analysis is shown in appendix \ref{appendix}. It
has to be pointed out here that a Floquet analysis has been applied to
our equations to investigate the nature of the bifurcation undergone
by the system at that threshold. We find that two purely imaginary conjugate Floquet multipliers (reflecting the behaviour of a perturbation to the
periodic solution after one cycle) cross the unit circle, thus
indicating a Hopf bifurcation.

\begin{figure}[h!]
	\centering
	\includegraphics[width=9.cm]{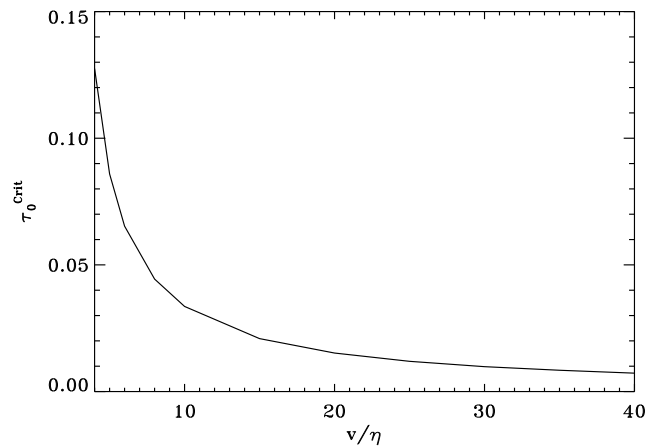}
	\caption{Value of $\tau_0$ for which the harmonic solution
          becomes unstable with respect to the normalized meridional
          flow speed.}
	\label{figure_tauvsv}
\end{figure}

Figure \ref{figure_tauvsv} shows the dependence of the critical
$\tau_0$ (i.e. the value for which the harmonic solution becomes
unstable) with the meridional flow speed normalized by the magnetic
diffusivity. In our particular case where $k=1$ (and thus where
  $v/\eta$ represents the ratio between diffusive and advective
  times), this gives us some insight on the value of the delay
which has to be reached to get a modulation, when the system tends
more and more towards an advection dominated regime. The trend of the
threshold to decrease with increasing meridional flow speed is not
surprising if we consider we are introducing fixed values for the
delays, not normalized to the cycle period. Indeed, as $v$ is
increased, the cycle period is decreased and then the absolute value
for the delay to have a significant effect on the cycle is
decreased. What is more striking is the saturation that the curve seems to
reach when $v$ is increased to strong values. In this case, the cycle
period {\bf (which is roughly inversely proportional to the meridional
  flow speed in this type of models)} is still decreased but the critical value for the delay seems
to reach a minimum value. We thus conclude that the percentage of the
cycle represented by the critical delay will increase when the meridional flow
speed is increased. As an example, in the case where the meridional
flow speed is quite weak ($v/\eta=4$), the threshold delay represents
about $14\%$ of the cycle period whereas at $v/\eta=40$, this
percentage reaches $30\%$. In this strongly advection-dominated
regime, the meridional flow speed is probably dominating the
time-scale of the system and as a result the effect of the delays is
reduced. 

In the remainder of the paper, we focus on an intermediate case
($v/\eta=10$), where the threshold value for the delay
($\tau_0\approx3.36\times 10^{-2}$) reaches $19\%$ of the cycle. This value
can appear quite large to the typical values for the delay used in the
full 2D model but it has to be pointed out that as soon as the
modulation appears, the strongest fields reach higher values than the
peak values of the
harmonic solution. As a consequence, the delays for the strongest
fields are shorter and reach about $8\%$ of the cycle, which is closer
to what was used in the 2D model.
This remark is related to the effect of fixed versus varying delays,
which we discuss in the conclusion.

%Influence of the parameters, especially meridional flow.

\subsection{Further evolution}

Knowing the behaviour of the harmonic solution (which is stable for
small values of the delay) and the influence of the parameters,
we choose a particular set of those parameters to try to study a generic
solution when the delay is further increased. We thus choose numbers which
satisfy the conditions invoked in section \ref{6thorder} and a
meridional flow speed in agreement with what was shown in the previous
section. The parameters chosen for all the analysis are the following:
$\Omega=100$, $S=100$,
$\eta=10^{-3}$, $v=10^{-2}$, $\lambda=10^{-4}$, the same as the ones
used for the solution shown on Figure \ref{figure_amp}.

\begin{figure}[h!]
	\centering
	\includegraphics[width=8.5cm]{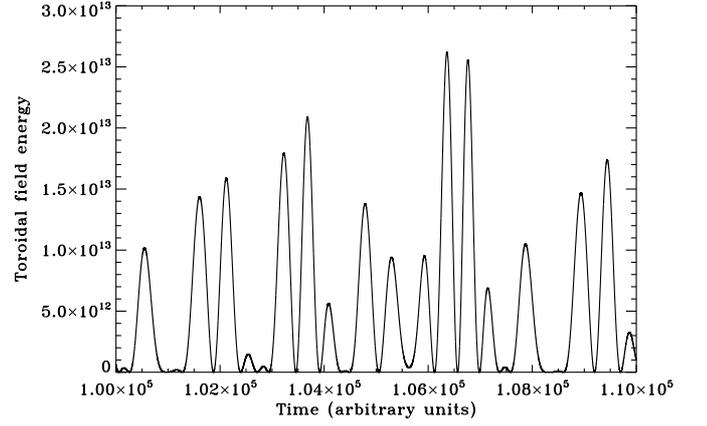}
	\caption{Time evolution of the toroidal magnetic energy with a
        value of the delay of $\tau_0=9\times10^{-2}$. Aperiodic
        modulation of the cyclic pattern is now fully developed.}
	\label{figure_9d-1}
\end{figure}

Figure \ref{figure_9d-1} shows the time series of the toroidal
magnetic energy for the solutions of the system when the delay is
  further increased to $\tau_0\approx9\times10^{-2}$. An aperiodic modulation of the solar cycle
is here clearly visible, and the major point is that the evolution of
toroidal energy is very similar to what was found in the full 2D
system when the delay was sufficiently large (see
Fig. \ref{figure_5e4}). Indeed, the major modulation acts on the cycle
amplitude but the frequency also varies and periods of very low
activity arise. This type of behaviour is very attractive when
compared to the observed sunspot number which qualitatively shows the
same kind of long-term variations. We should stress that we are still
here in a regime where the time delays are small compared to the
long-term modulation they seem to create. Moreover, this perturbed solution
is obtained both through 2D mean-field simulations and through a
reduced set of ODEs solving the Babcock-Leighton dynamo problem. These
results are
thus of particular interest as a new origin of modulation seems to
appear: the time delays due to the buoyant rise of flux tubes from the
base of the convection zone to the solar surface to create active
regions.
We now need to understand what happens to the system between the first
bifurcation and this aperiodic more interesting solution.

We thus turn to
investigate in more details which sequence of bifurcations lead to such
complicated behaviour.

\section{Beyond the first bifurcation: reduction to a 5th order
  system}
\label{sect_5th}

To gain some insight on the behaviour of the system after the first
bifurcation, we show that we can further reduce our problem from a 6th
order to a 5th order system. Indeed, it is rather impractical to study
the precise behaviour of the system after the first bifurcation. Reducing our system even more will help
us achieve this goal.

\subsection{Formulation of the model}

Noticing that our system possesses a phase-invariance, i.e. a further
symmetry, we follow the procedure of Weiss et al. (1984) and \cite{Jones85} and make a change
of variables which will remove this symmetry.

We express our variables $A_t$, $B_t$ and $Q_t$ using polar
coordinates and introduce 
four new variables as shown

$$
A_t=\rho \, y \, e^{i\theta}
$$
$$
B_t=\rho \, e^{i\theta}
$$
$$
Q_t=\rho \, z \, e^{i\theta}
$$

\noindent where y and z are complex numbers and $\theta$ and $\rho$
are real. The real numbers $\rho$ and $\theta$ here represent
respectively the modulus and argument of the basic toroidal field
$B_t$. We can now reintroduce these expressions in equations
\ref{eqA}, \ref{eqB} and \ref{eqQ} to get, after some algebra, the new set of ODEs

\begin{equation}
\dot{\rho}=-\, \Omega \, \rho \, \Im(y) - \, \eta \, \rho
\label{eqrho}
\end{equation}
\begin{equation}
\dot{y}=\frac{S \, z}{1+\lambda \, \rho^2 \, \vert z \vert^2}- i \, \Omega \, y^2
\label{eqy}
\end{equation}
\begin{equation}
\dot{z}=\frac{1-z}{\tau}-(i \, \Omega \, y - i \, v - \eta) \, z
\label{eqz}
\end{equation}

\noindent where
$$
\tau=\frac{\tau_0}{1+\rho^2}
$$

We then get a 5th order system since $y$ and $z$ are
complex but $\rho$ is real. We find that $\theta$, which is the phase
of the ``undelayed'' toroidal field, does not appear
explicitly in this new set of equations but can be obtained from

\begin{equation}
\dot{\theta}= \Omega \, \rho \, \Re(y)- v \, \rho
\label{eqtheta}
\end{equation}

%\begin{figure}[h!]
%	\centering
%	\includegraphics[width=9.cm]{./phase_3_32_5th.ps}
%	\includegraphics[width=8.cm]{./bphi30_lag5e4.ps}
%	\caption{Phase portrait in the  $(\Re(\rho \, y), \rho)$ plane of the solution of the 5th order model
%        after the first bifurcation (at $\tau_0=3.36 \times
%        10^{-2}$). We now have a closed curve in the phase space,
%        corresponding to a periodic orbit for the parameter for which
%        we had the torus of Fig. \ref{figure_hopf}.}
%	\label{figure_hopf_5th}
%\end{figure}

The interesting outcome of reducing our system from 6 degrees of
freedom to 5 is that the limit cycle of the higher-order system now
corresponds to a stationary solution in our reduced model and that a
torus now corresponds to a limit cycle. 
%As a consequence, the time
%evolution of $\rho$, the modulus of $B_t$ switches from a flat curve
%before the Hopf bifurcation exhibited in the previous section to a
%periodic evolution. 

%Figure \ref{figure_hopf_5th}
%shows one phase portrait of the solution to the new set of ODEs for
%the same value of the parameter $\tau_0$ as in
%Fig. \ref{figure_hopf}, i.e. just after the first bifurcation. The
%phase portrait represented here in a particular plane is then no
%longer a torus but a closed curve characteristic of the presence of a
%stable limit cycle. Our new
%system has thus bifurcated from a stationary solution to a periodic one and
%we note that $\rho$, representing the modulus of the toroidal field
%$B_t$, has indeed the same values as in the higher-order system.

\begin{figure*}[htbp]
	\centering
	\includegraphics[width=18cm]{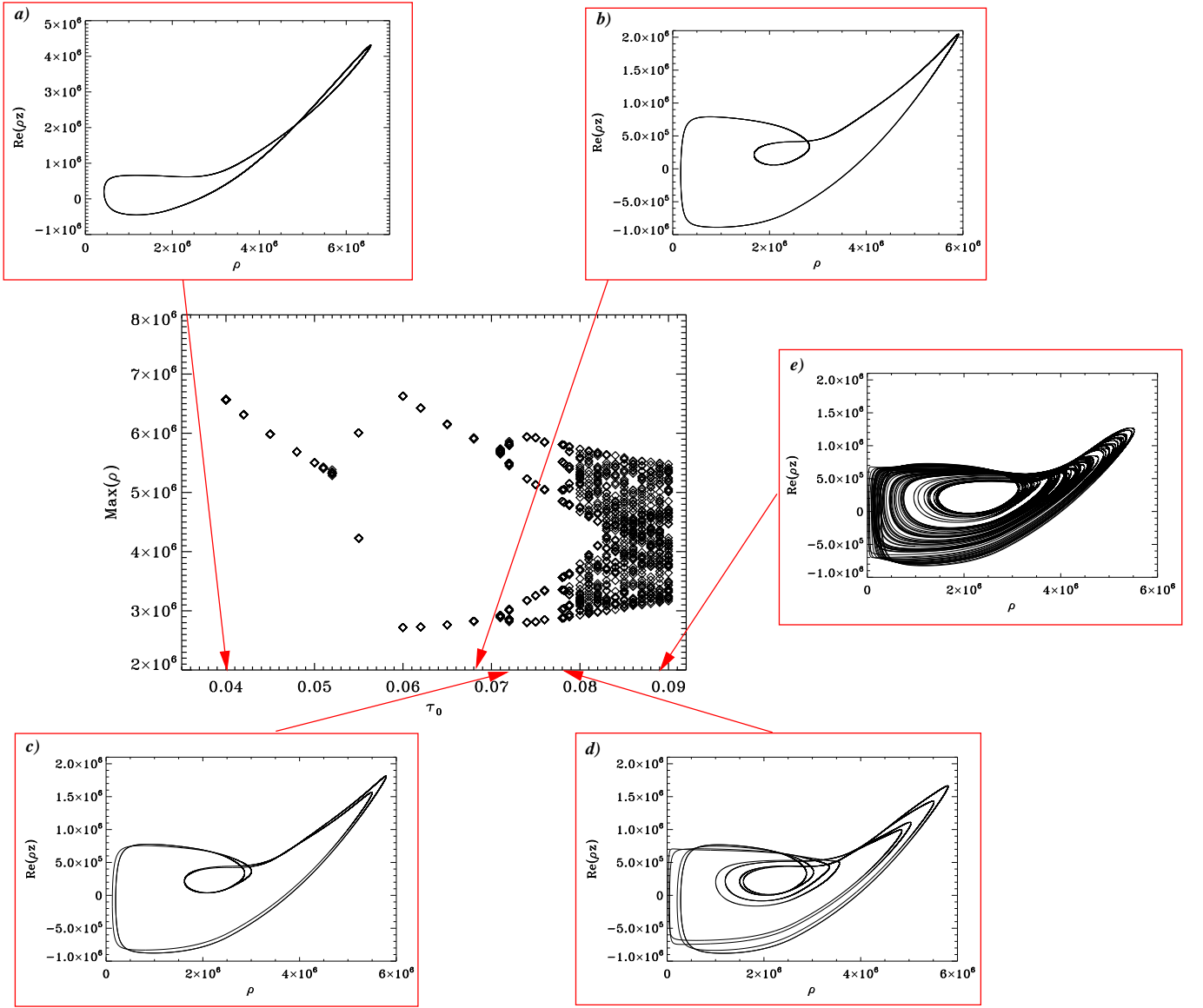}
	\caption{Evolution of the system after $\tau_0=4\times10^{-2}$: bifurcation diagram and phase portraits in the
          $(\rho,\Re(\rho \, z))$ plane for values of the
          control parameter $\tau_0$ close to those for which successive period doublings occur.}
	\label{figure_bif}
\end{figure*}

Solving this new system numerically will now enable us to
analyze the further evolution of our solution when the delay is
increased. In particular, we show that we can now easily exhibit the sequence of bifurcations
leading to the aperiodic modulation of Fig. \ref{figure_9d-1}.

\subsection{From the harmonic to the aperiodic solution: a sequence of
bifurcations}

 We follow the evolution of our system from the first bifurcation
 at $\tau_0=3.36\times10^{-2}$ to the aperiodic solution at $\tau_0=9
 \times 10^{-2}$. First, we can relate again these values to some physically
 meaningful time-delays, keeping in mind that the system has been
 simplified and that a particular choice of parameters was made. For
 the stronger fields, a
 value of $\tau_0 \approx 3.36\times10^{-2}$ would then correspond to a delay
 of approximately a tenth of the cycle duration and
 $\tau_0 \approx 9\times10^{-2}$ a third of the cycle period. These delays may
 look higher than what was calculated in the 2D model but we showed
 that a modification of the parameters (and especially of the
 meridional flow speed) could move the threshold above which the
 modulation starts to appear.

%Since the 2D model is strongly dominated
% by advection by the meridional flow, the appearance of a modulation
% at smaller delays is then consistent with our results.

%Figure \ref{figure}
 The Floquet
multipliers for this system are computed in order to study the stability of our periodic solution.
We find that the solution of our new reduced model
stays on a periodic orbit for a large range of values for the delay,
i.e. from $3.36\times10^{-2}$ to about $5.43\times10^{-2}$. We
indeed find that at this value for the parameter $\tau_0$, one of the
real Floquet multipliers crosses the unit circle through the -1 point,
indicating a period doubling bifurcation. This new 2-periodic solution then
persists for quite a large range of the control parameter until
we reach a new period doubling bifurcation.

Figure \ref{figure_bif} shows the evolution of the system from the
periodic solution resulting from the first Hopf bifurcation at
$\tau_0 \approx 3.36\times10^{-2}$ until we reach the aperiodic modulations of
the cycle. We present on this figure a bifurcation diagram computed by
determining the maximum values of $\rho$ for each value of $\tau_0 \geq
4\times10^{-2}$ (when the first period doubling has already
occured). The initial branch existing at the beginning of the
bifurcation diagram thus indicates a periodic orbit. Moreover, phase portraits at key values of
the parameter are shown. They were chosen to stress the sequence of
period doubling bifurcations exhibited by the set of ODEs. %Indeed, the
%solution initially follows a regular evolution, shown on panel a) of
%Fig. \ref{figure_bif}. 
Panel b) shows the modification of the phase portrait of panel a) after the first period doubling has
occured. %We note that the period doubling is responsible for the
%particular shape of this phase portrait, especially the loop showing
%in the centre of the figure. 
We are then in the presence of a 2-periodic solution:
every 2 cycles, the value of $\rho$ reaches the
same maximum value and this behaviour persists for as long as the model
has been calculated, i.e. for thousands of cycles. At $\tau_0 \approx 7.15
\times 10^{-2}$, the system goes through another bifurcation. The
period is doubled again, which is visible both on the bifurcation
diagram (where the number of points is doubled at this value) and on
panel c). Indeed, the curve on the phase portrait closes on
itself after 4 cycles now instead of 2 for the previous values of
$\tau_0$. Period doublings happen again at $\tau_0 \approx 7.8
\times 10^{-2}$ and  $\tau_0 \approx 7.88
\times 10^{-2}$ where the solution becomes first 8-periodic and then
16-periodic. The phase portrait on panel d) shows the time
needed to close the curve is doubled at $\tau_0 \approx 7.8
\times 10^{-2}$ compared to the solution of panel c). %It is especially obvious
%on the upper right part of this panel, where the curve is
%clearly doubled compared to the previous phase portrait. 
But
obviously, since the curve is
still closed, the system is still periodic.

 It then becomes difficult
to identify further period doublings but they may occur in very narrow
windows for the parameter. The main point is that the system seems to
reach a very complicated behaviour when the value of $\tau_0$
approaches $8.5 \times 10^{-2}$. Panel e) shows the phase portrait for
$\tau_0=8.8 \times 10^{-2}$. The phase space is almost entirely covered
by the solution and any value can be reached by the peaks in the
temporal evolution of $\rho$, as shown on the bifurcation diagram. At this stage, a clear periodicity is
difficult to identify and we recover the strongly modulated solution
shown on Fig. \ref{figure_9d-1} for a value of the parameter of $9\times10^{-2}$.

%\bigskip

Consequently, our system seems to undergo a sequence of period doublings
leading to a chaotic behaviour. The solution may well be modified
for higher values of the time-delays but which would not be realistic anymore for the
physical process we are modelling. Similar cascades of period doublings
have been found in the Lorenz equations (\cite{Sparrow82}) and in other systems of
ordinary and partial differential equations, including
those governing the solar dynamo (\cite{Moore83, Knobloch83, Weiss84,
  Knobloch98}). However, the nonlinearities at the origin of the
sequence of bifurcations in those models were always related to the
dynamical feedback of the Lorentz force on the flow. The striking
result in this work is that modulation of the cycle amplitude can also
arise in models where the only nonlinearities are the quenching term in
the Babcock-Leighton source and the magnetic field-dependent time
delays. It has here to be noted though that the time delays also
appear in the quenching term (see Eq. \ref{eq_s}), possibly amplifying the nonlinear
effects in the model.  

%\bigskip

We conclude that time delays due to flux tubes rising from the base of the
convection zone to the surface seem to create a strongly modulated
cycle in Babcock-Leighton dynamo models. A sequence of period
doublings leading to chaos has clearly been identified here thanks to
our 5th-order system. Indeed, even though it is also obviously
present in the 6th order system, it is much more difficult to exhibit. The magnetic cycle
resulting from these models which take into account the rise time of
toroidal fields thus seems to be closer to the
actual behaviour of the solar cycle. 

%\newpage

\section{Discussion and conclusion}
\label{sect_conclu}

In this work, we have introduced a more physically accurate
model for the poloidal field source term in Babcock-Leighton flux
transport dynamo models. To do so, some results from 3D MHD calculations of
rising flux tubes from the base of the convection zone to the surface
were reintroduced in 2D mean-field models. In particular, the rise
time of flux ropes, which was assumed to be negligible in previous
simulations, was here taken into account by introducing magnetic
energy-dependent time delays in the BL source term. Through the use of
a combination of analytical and numerical techniques applied to an
adapted reduced system of non linear equations, we show that the
system exhibits a sequence of bifurcations leading to a chaotic
behaviour. The time evolution of the solution at that stage is
qualitatively very similar to the full 2D model and more importantly to the actual solar activity, with strong
modulation especially on the cycle amplitude.

\begin{figure}[h!]
	\centering
	\includegraphics[width=8.5cm]{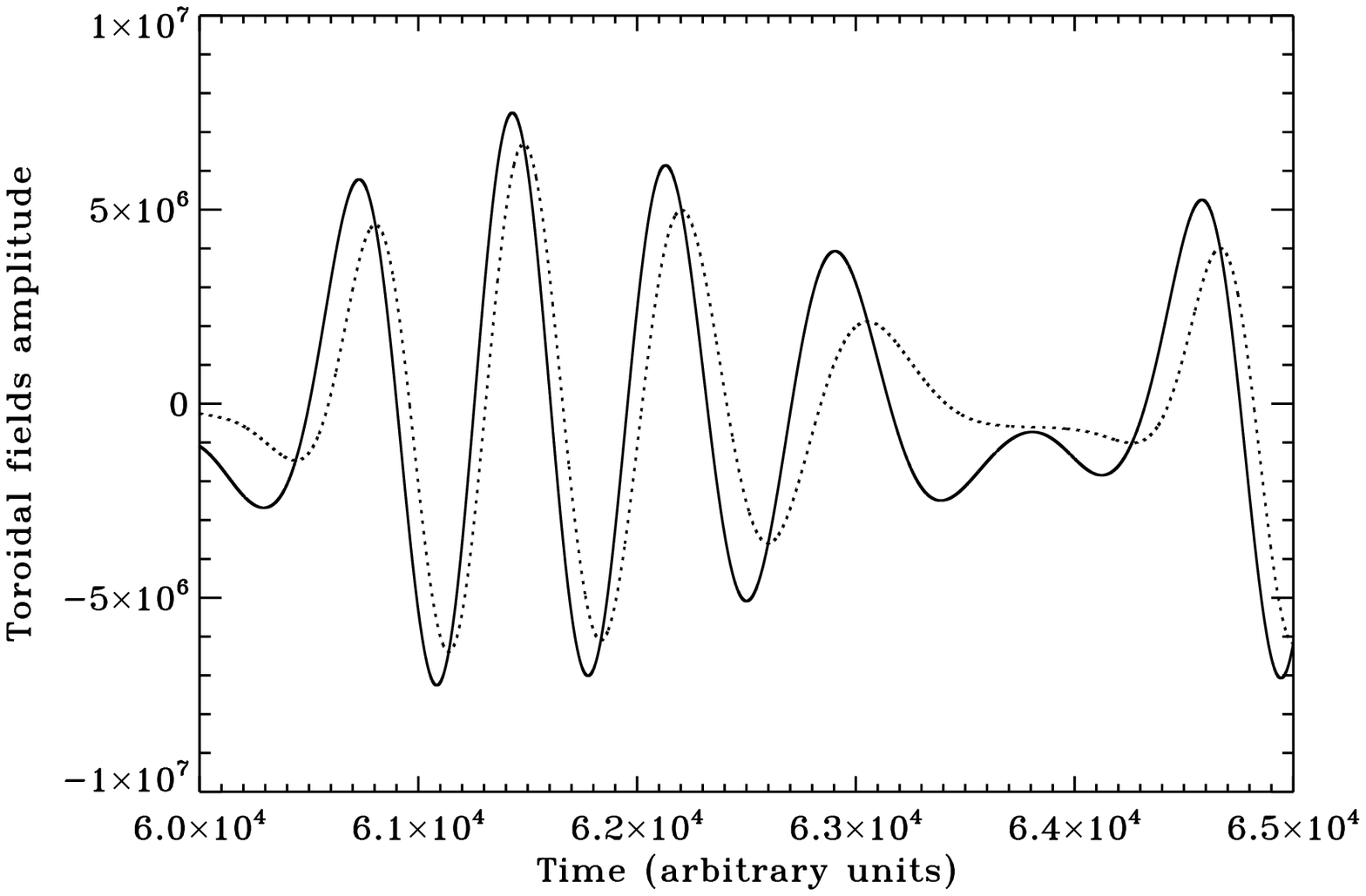}
	\includegraphics[width=8.5cm]{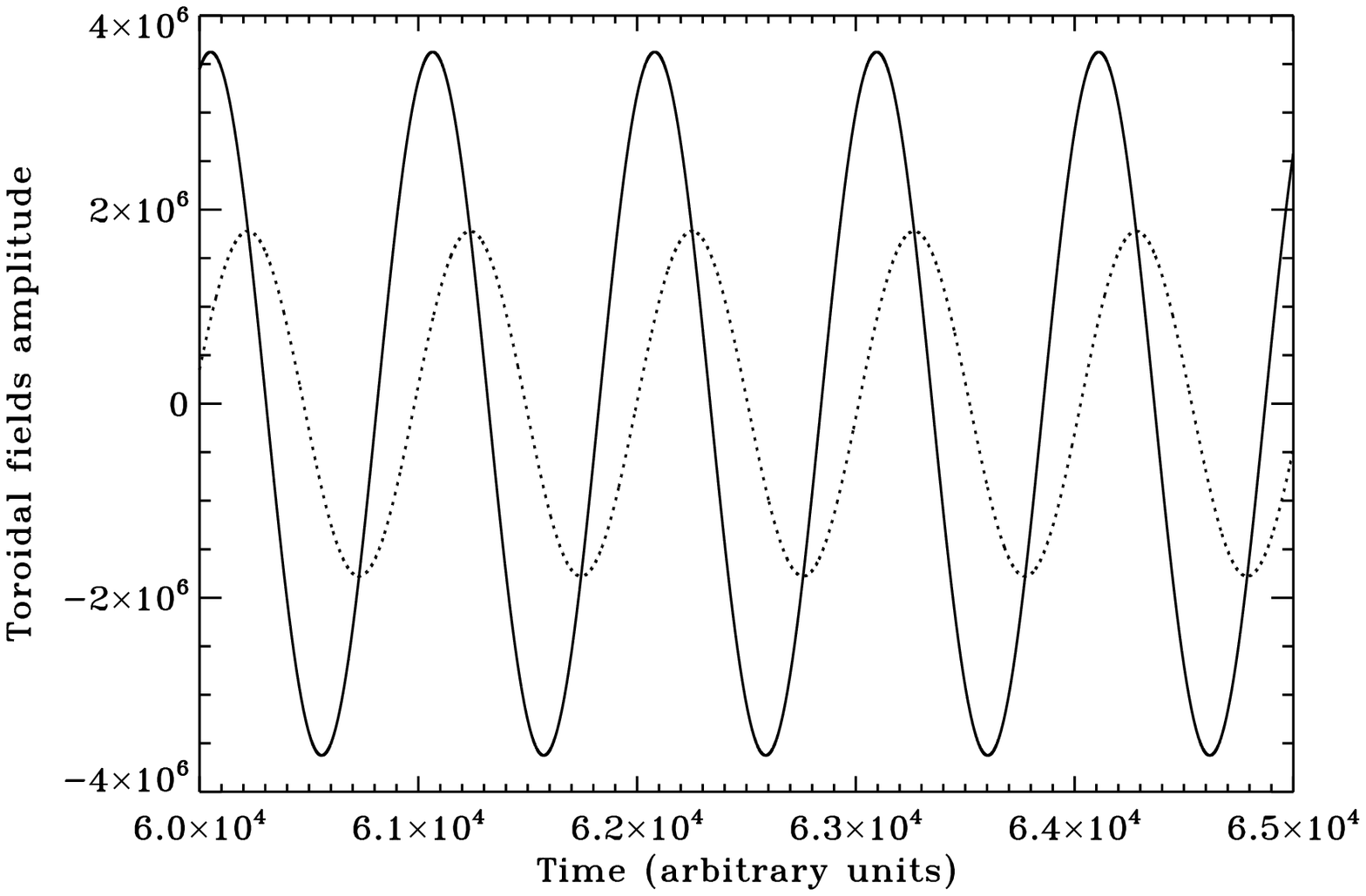}
	\caption{Time evolution of the toroidal fields amplitude (undelayed in
          plain line and delayed in dotted line) for a
          delay dependent on $\vert B \vert^2$ just after the Hopf
          bifurcation (upper panel) and for a delay fixed
          to the mean value of the previous run (lower panel).}
	\label{figure_bq_mod}
\end{figure}

Assuming the rise of flux tubes from the base of the convection zone
and the surface to be instantaneous and thus the time delays due to magnetic
buoyancy to be negligible can sound reasonable since these delays are
weak compared to the other time-scales of the system. In particular,
we can expect the time delay due to the advection by the meridional
flow (which has a characteristic time-scale of a few solar cycles, see
for example \cite{Charbonneau00}) to
be much more effective on the behaviour of the dynamo-generated
magnetic field. As a consequence, it is mainly the effect of
meridional flow which has been tested in previous models (e.g. \cite{Wilmot05}). However, we showed here that the time delays due to
flux tubes rising have in fact a significant effect to modulate the
cycle on time scales very large compared to the delays. The main
reason for this behaviour is that the delays are not fixed but depend
on the modulated toroidal energy itself. To illustrate this
explanation, we show on Figure \ref{figure_bq_mod} the time evolution of
toroidal fields (delayed and undelayed) for a case similar to what we showed in the core of the
paper (a $\vert B \vert^2$-dependent delay with $\tau_0=3.4 \times
10^{-2}$, i.e. just after the Hopf bifurcation) and the same evolution
when the delay is fixed to the mean value reached by the delay in the previous run.

%\subsection{A new origin of the solar cycle modulation?}

On this figure, both the basic toroidal field and the delayed field
are shown, stressing the time shift existing between the two.
The striking result is that a long-term modulation appears only in the
case of a varying delay. In the fixed-delay case, although the
time-shift between the basic toroidal field and its delayed
counterpart is quite significant compared to the cycle period, no
modulation is created. On the contrary, in the varying-delay case,
even though the strongest fields are almost not affected by the
time-shift, a strong modulation on the amplitude is built up. 
As a result, the fact that small delays can affect the long-term
evolution of the dynamo cycle seems to be linked to the variability of
these delays and more specifically to their dependence on the magnetic
field strength. Since magnetic buoyancy is much more efficient in the
solar interior for strong toroidal field structures, these kind of
delays are likely to appear in reality. Consequently, a new origin of
the solar cycle modulation may have been revealed here, thanks to the
input of 3D MHD models into 2D mean-field dynamo calculations. 

Only a part of the results of 3D calculations were reintroduced here
but other effects can also be taken into account. In particular, hoop
stresses and the Coriolis force are known since the first thin flux
tubes simulations to deflect the trajectory of buoyantly rising
magnetic fields inside the convection zone, this effect being stronger
when the fields are less intense. This feature could also be introduced
in BL models by modifying the source term for poloidal field. In
particular, the source term is likely to get a contribution from the
same latitude a the initial position of the flux tubes when those are
strong enough and from a lower latitude when the flux tubes have been
sufficiently influenced by the Coriolis force and the hoop stresses to
rise parallel to the rotation axis. This may result in a different
shape for the butterfly diagram with slightly more activity at higher
latitudes, provided that weak magnetic structures are assumed to be
able to reach the surface and create active regions. This could lead
to reconsider the correspondence which is often made between toroidal magnetic field generated in 2D dynamo
models at the base of the convection zone and the actual sunspot
migration observed at the solar surface.

The variations of the solar cycle period are also known to be
significant. For instance, cycle 23 has been significantly longer
than the previous ones, with a duration of about 13 years
(1996-2009). In the models computed in this work, the modulation of
the solar activity is particularly visible on the cycle amplitude and
is much less obvious on the period. However, the influence of rising
flux tubes on the meridional flow was not taken into account
here. Since the cycle period of the solutions resulting from these
flux-transport models is very sensitive to the amplitude of the
meridional circulation, a modification of the flow by rising flux
tubes is very likely to modify the frequency of the dynamo solution
and thus to also introduce a modulation on the cycle period. This
feature is currently being investigated.  

A considerable step forward would obviously be to develop a  
  self-consistent global model with buoyant toroidal structures built
  up and making their way from the base of the convection zone where
  they become unstable to the photosphere where they create active
  regions. Unfortunately, this has not been achieved yet due to numerous
  physical and numerical difficulties. 
In the mean time, we have shown here that inputs from 3D MHD models simulating
a particular step of the dynamo cycle can significantly
improve 2D mean-field calculations. In particular, we have shown that
they can even help reproducing a feature of the solar cycle (its
variability) which was not present in the standard model before. We
will continue to develop these ideas in future work.

\begin{acknowledgements}

LJ and GL acknowledge support
   by STFC.

\end{acknowledgements}

%
%\newpage%%%%%%%%%%%%%%%%%%%%%%%%%%%%%%%%%%%%%%%%%%%%%%%%%%%%%%

\begin{appendix}

\section{Amplitude and frequency of the harmonic solution}
\label{freq}

We can
further expand our harmonic solution on a single Fourier mode in time as:

$$
A_t(t)=A_0 \exp(i \omega t) 
$$
$$
B_t(t)=B_0 \exp(i \omega t)
$$
$$
Q_t(t)=Q_0 \exp(i \omega t)
$$

Reintroducing these expansions in equations \ref{eqA}, \ref{eqB} and
\ref{eqQ} will give us a way to calculate the frequency $\omega$ and the
amplitude of the harmonic solution $\vert B_0 \vert$ or equivalently
$\vert Q_0 \vert$ with respect to all the
parameters. The complex dispersion relation we get with $k=1$ is the following

\begin{equation}
(i \, \omega +i \, v + \eta)^2 \, (i \, \omega \, \tau+1)=\frac{i \, \Omega \, S}{1+\lambda
  \, \vert Q_0 \vert^2}
\label{eq_disp}
\end{equation}

Taking the real and imaginary parts of this relation dispersion give
us the following equations

\begin{equation}
\omega^2 (1+2 \, \eta \, \tau) + 2 \omega \, v (1+\tau \, \eta) + v^2
- \eta^2 =0
\end{equation}

\begin{equation}
\omega^3 \, \tau + 2 \omega^2 \, v \, \tau + \omega (v^2 \, \tau -
\eta^2 \, \tau - 2 \eta) =- \frac{\Omega \, S}{1+\lambda
  \, \vert Q_0 \vert^2}+2 \eta \, v
\end{equation}

\noindent where 
$$
\tau=\tau_0/(1+\vert B_0 \vert^2)
$$

\section{Stability of the harmonic solution}
\label{appendix}

We can focus on the stability of this
periodic solution by perturbing it and studying the growth rates of
the perturbations. To do so, we perturb the solution as follows

\begin{equation}
\tilde{A}=A_t \, (1+\alpha_1 \, e^{pt}+ \alpha_2^{\star} \, e^{p^{\star}t})
\end{equation}

\begin{equation}
\tilde{B}=B_t \, (1+\beta_1 \, e^{pt}+ \beta_2^{\star} \, e^{p^{\star}t})
\end{equation}

\begin{equation}
\tilde{Q}=Q_t \, (1+\gamma_1 \, e^{pt}+ \gamma_2^{\star} \, e^{p^{\star}t})
\end{equation}

\noindent where $\alpha_1$, $\alpha_2$, $\beta_1$, $\beta_2$,
$\gamma_1$ and $\gamma_2$ represent the coefficients of the perturbed
fields, $p$ represents the complex growth rate of the perturbation and
the symbol $\star$ stands for the complex conjugate.

Substituting these expressions in equations \ref{eqA}, \ref{eqB} and
\ref{eqQ} give us a system of 6 equations relating the coefficients of
the perturbations, the growth rate and all the parameters of the
system. 

$$
(i\omega+p+\eta+iv)\, \alpha_1=(i\omega+\eta+iv)\big[\gamma_1-\frac{\lambda
  \vert Q_0 \vert^2}{1+\lambda \vert Q_0 \vert ^2} \,(\gamma_1+\gamma_2)\big]
$$
$$
(-i\omega+p+\eta-iv)\,\alpha_2=-(i\omega-\eta+iv)\big[\gamma_2-\frac{\lambda
  \vert Q_0 \vert^2}{1+\lambda \vert Q_0 \vert ^2} \,(\gamma_1+\gamma_2)\big]
$$
$$
(i\omega+p+\eta+iv)\,\beta_1=(i\omega+\eta+iv)\,\alpha_1
$$
$$
(-i\omega+p+\eta-iv)\,\beta_2=(-i\omega+\eta-iv)\,\alpha_2
$$
$$
(i\omega+p+\frac{1+\vert B_0
  \vert^2}{\tau_0})\,\gamma_1=(i\omega+\frac{1+\vert B_0
  \vert^2}{\tau_0}+\frac{i \omega \vert B_0 \vert^2}{1+\vert B_0 \vert^2})\beta_1+\frac{i\omega\vert
B_0 \vert^2}{1+\vert B_0 \vert^2}\,\beta_2
$$
$$
(i\omega-p-\frac{1+\vert B_0
  \vert^2}{\tau_0})\,\gamma_2=(i\omega-\frac{1+\vert B_0
  \vert^2}{\tau_0}+\frac{i \omega \vert B_0 \vert^2}{1+\vert B_0 \vert^2})\beta_2+\frac{i\omega\vert
B_0 \vert^2}{1+\vert B_0 \vert^2}\,\beta_1
$$

\noindent $\vert B_0 \vert^2$ and $\vert Q_0 \vert^2$ being related by
the following expression

$$
\vert Q_0 \vert^2=\frac{\vert B_0 \vert^2 (1+\vert B_0
  \vert^2)^2}{\omega^2\tau_0^2+(1+\vert B_0 \vert ^2)^2}
$$

A non-trivial solution for this system exists only if the determinant
is zero, we thus have to calculate the determinant and find the roots
of the 6th order polynomial, in terms of the growth rate $p$. The
harmonic solution loses stability at a Hopf bifurcation when the real
part of one the roots becomes positive.

\end{appendix}

\end{document}